\documentclass[review,times,a4paper]{elsarticle}

\usepackage[USenglish]{babel}
\usepackage[utf8]{inputenc}
\usepackage[T1]{fontenc}

\usepackage{natbib}
\setcitestyle{authoryear}

\usepackage{amsmath}
\usepackage{subcaption}
\usepackage{booktabs}
\usepackage{amssymb}
\usepackage{graphicx}
\usepackage{caption}
\usepackage{multirow}
\usepackage{color,soul}
\usepackage[super]{nth}
\usepackage{graphicx}
\usepackage{algorithm}
\usepackage{siunitx}
\usepackage{float}
\usepackage{url}
\usepackage{ulem}
\usepackage[noend]{algpseudocode}
\usepackage[capitalize]{cleveref}
\usepackage{xcolor}
\usepackage[toc,page]{appendix}
\usepackage[version=4,arrows=pgf-filled,
textfontname=sffamily,
mathfontname=mathsf]{mhchem}

\graphicspath{{./figures/}}

\def\eps{\varepsilon}
\def\sig{\sigma}

\def\stress{\boldsymbol{\sigma}}

\def\disp{u}

\def\biot{\alpha}

\def\testv{s}
\def\sif1{K_I}

\def\open{\delta_n}
\def\slip{\delta_s}
\def\co2{\textrm{CO}_2}
\newcommand{\av}[1] {\langle {#1} \rangle}
\newcommand{\jump}[1] {[[ {#1} ]]}

\begin{document}

\begin{frontmatter}
  \title{A Parallelized 3D Geomechanical Solver for Fluid-induced Fault Slip in Poroelastic Media}
  
  \author[a]{E.R.~Gallyamov}
  \author[a]{G.~Anciaux}
  \author[a] {N.~Richart}
  \author[a,b]{J.-F.~Molinari}
  \author[a]{B.~Lecampion\corref{cor1}}\ead{brice.lecampion@epfl.ch}
  \cortext[cor1]{Corresponding author}

  \address[a]{Civil Engineering Institute, École Polytechnique Fédérale de Lausanne
    (EPFL), Station 18, CH-1015 Lausanne, Switzerland}
  \address[b]{Materials Science and
    Engineering Institute, École Polytechnique Fédérale de Lausanne
    (EPFL), Station 18, CH-1015 Lausanne, Switzerland}
    
  \begin{abstract}
    We present a fully implicit formulation of coupled fluid flow and geomechanics for fluid injection/withdrawal in fractured reservoirs in the context of $\co2$ storage.
    Utilizing a Galerkin finite-element approach, both flow and poroelasticity equations are discretized on a shared three-dimensional mesh. The fluid flow is assumed to be single-phase. The hydraulic behaviour of fractures is represented through a double-nodes flow element, which allows to efficiently model longitudinal and transversal fracture permeabilities. In addressing the mechanical subproblem, fractures are explicitly modelled using cohesive elements to account for contact, friction and opening phenomena. The nonlinear set of equations is solved implicitly through an iterative partitioned conjugate gradient procedure, extending its traditional application to continuous problems to those involving explicit  discontinuities such as faults and fractures. The model's accuracy is verified against 
    analytical solutions for different geomechanical problems, 
    notably for the growth of a frictional slip rupture along a fault due to fluid injection. Such a particularly challenging benchmark for a critically stressed fault is here reproduced for the first time by a finite-element based scheme.    
    The capabilities of the developed parallel solver are then illustrated through a scenario involving $\co2$ injection into a faulted aquifer. The original solver code, tutorials, and data visualization routines are publicly accessible.
  \end{abstract}
  
  \begin{keyword}
    geomechanics \sep $\co2$ storage 
    \sep finite elements \sep fluid-induced fault slip
  \end{keyword}
\end{frontmatter}


\section{Introduction}
Geologic carbon storage (GCS) involves injecting carbon dioxide into underground geological formations to prevent its release into the atmosphere. Recent examples of GCS demonstrate that this technology is not only feasible but also effective in mitigating the impacts of anthropogenic $\co2$ emissions \citep{maldal_co2_2004,ringrose_salah_2013}. The captured $\co2$ is used predominantly for enhanced oil recover (EOR), as in the case of Boundary Dam and Petra Nova, USA \citep{mantripragada_boundary_2019}, the Weyburn Project, Canada \citep{preston_iea_2005}, or for direct geological storage, as demonstrated by the Gorgon Project, Australia \citep{trupp_developing_2021} and the Sleipner Project, North Sea \citep{torp_demonstrating_2004}.

As with any underground anthropogenic activity,  certain technical risks \citep{anderson_risk_2017} must be managed in relation to GCS. In particular, the integrity of the wells and the reservoir seals must be ensured during and decades after the injection.
Among the most detrimental geomechanical risks are induced ground motion, exemplified by surface uplift observed in the In Salah gas field, Algeria \citep{vasco_satellite-based_2010,lecampion_inversion_2011}; induced seismicity, reported in various locations in the USA, Canada, and Algeria \citep{white_assessing_2016}; and the leakage of $\co2$ and fluid displacement out of storage reservoirs \citep{birkholzer_co2_2015,pawar_recent_2015}.
To mitigate these risks and enhance operational efficiency and predictability, substantial scientific research has been devoted to studying $\co2$ injection from both the experimental \citep{ali_recent_2022,peter_review_2022} and modelling \citep{liu_co2_2022, kalam_carbon_2021} perspectives. 
Modelling is an important part of the engineering of storage sites within a classical Monitoring-Modelling-Verification (MMV) workflow. The pioneering work in numerical modelling of geomechanical problems includes studies by \cite{sandhu_finite-element_1969,christian_undrained_1968,booker_investigation_1975}. Additionally, the textbook by \cite{lewis_finite_2000} remains a primary reference in computational geomechanics.

The approaches to geomechanical modelling of fractured reservoirs can be grouped into two high-level categories: continuum and discrete models. The first type of models does not explicitly represent fractures, but their effect on flow and deformations is accounted for by using effective or upscaled flow and geomechanical properties. In discrete models, the geomechanical processes are simulated with an explicit representation of the rock matrix, fractures and faults. The most common continuum approaches are the equivalent porous medium (EPM), stochastic continuum (SC), and dual continuum models (DCM). In EPM approaches, properties of fractures, faults, and their networks are homogenized to define continuum-scale effective properties that characterize the joint behaviour of the rock and fractures. Typical examples of such models are found in \cite{botros_mapping_2008,hartley_approaches_2013,sweeney_upscaled_2020}. The definition of the hydromechanical properties in EPMs is deterministic and depends on the specific realization of a fracture network. When information about specific fracture locations is scarce, the effective properties of fractured rock can be determined based on stochastic properties of the underlying fracture network, which form the essence of stochastic continuum approach (SC) (see \cite{neuman_trends_2005} for examples). Dual continuum models (DCM) establish two separate domains with unique properties, one representing the rock matrix continuum, and another the fracture network, with fluxes of fluid and energy between them being represented by linear exchange formulations. Examples of such models can be found in \cite{ashworth_foundations_2019,ahmed_geomechanical_2020,liu_coupled_2013,moinfar_development_2013}.

For discrete approaches, the most common ones are the discrete fracture network (DFN) and the discrete fracture-matrix (DFM). DFN models explicitly resolve the geometry and properties of fractures, but do not directly represent the rock matrix. Such models are best suited for the applications where the rock matrix permeability is negligible, and most of the fluid flow and deformations are localized within the fracture network. Examples of DFN models used for geomechanical applications can be found in \cite{davy_model_2013,thomas_growth_2020,ciardo_injection-induced_2023,mcclure_fully_2016,meyer_discrete_2011}. Discrete fracture-matrix (DFM) models explicitly represent both the fracture network and the porous matrix. Such geomechanical models can be found in \cite{garipov_discrete_2016,gallyamov_discrete_2018,paluszny_hydro-mechanical_2020,conti_xfvm_2023,tripuraneni_nonlinear_2023,jha_coupled_2014}. A comprehensive review of models for flow in fractured porous media has recently been presented by \cite{berre_flow_2019}. Concurrently, \cite{viswanathan_fluid_2022} have thoroughly examined various modelling approaches that address hydro-thermo-chemo-mechanical coupling.

A typical geomechanical solver is based on the solution of fluid volume conservation equation for the fluid flow and the balance of momentum equation for the rock deformation, which will be discussed in the following section. The tangent linear system formed by the implicit-in-time discretization and linearization of these equations requires efficient solving methods. While direct solvers can handle small systems, iterative schemes are required for large problems. Traditionally, these schemes are classified into two distinct categories: fully-implicit (or "monolithic") and sequential-implicit methods (commonly referred to as "staggered"). Fully-implicit methods update the pressure and displacement unknowns simultaneously. For large-scale problems, they require Krylov-type iterative solvers along with effective preconditioners to address the ill-conditioning of the resulting system of equations. For most fully-implicit schemes, design of dedicated multiple-field solvers is required. However, \cite{prevost_partitioned_1997} introduced a fully-implicit approach that enables the use of existing single-field computational software to solve the coupled field problem. Detailed discussions on fully-implicit solvers can be found in the works of \cite{white_block-preconditioned_2011,prevost_two-way_2014,bergamaschi_novel_2007} among others.

In sequential-implicit schemes, the governing equations are solved independently, with updates to the displacement and pressure fields occurring alternately. During each update cycle, one field, or its derivative, is "frozen" while the other is modified. This iterative process continues until convergence is reached. Such iterative scheme allows separate use of specific single-field solvers for the momentum and mass balance sub-problems. The stability and convergence properties of various splitting schemes are comprehensively studied and discussed in detail in the works of \cite{kim_stability_2009,kim_stability_2011,girault_convergence_2016,mikelic_convergence_2013,white_block-partitioned_2016}. 


The accuracy of geomechanical solvers requires thorough verification. The most commonly employed verification tests for the poromechanics include several classic problems. The Terzaghi uniaxial consolidation problem \citep{terzaghi_erdbaumechanik_1925,cheng_poroelasticity_2016} serves as a fundamental model. Additionally, the Cryer saturated sphere \citep{cryer_comparison_1963, cheng_poroelasticity_2016} and the Mandel compressed 2D specimen \citep{mandel_consolidation_1953,verruijt_consolidation_2008} both demonstrate the Mandel-Cryer effect, which describes the non-monotonic fluid pressure variation over time which is an intrinsic feature of poroelasticity. Moreover, the Noordbergum effect, which involves an increase in fluid pressure due to fluid extraction from a hydraulically-isolated permeable layer, was first reported and explained by \cite{verruijt_elastic_1969}. A benchmark for frictional sliding of a dry single fracture in 2D is readily available in \cite{phan_symmetric-galerkin_2003}. 
However, it is only recently that analytical solutions have been obtained for the very important problem of fluid-induced frictional fault slip, which introduce the additional complexity of properly capturing the evolution of the moving slip front.
\cite{bhattacharya_fluid-induced_2019,viesca_self-similar_2021} have proposed analytical solutions for a 2D problem of injection-induced aseismic slip of a planar fault. 
\cite{saez_three-dimensional_2022} have investigated in details the three-dimensional case and obtained analytical solutions for circular fault-induced slip ruptures. 
These models, which integrate the dynamics of fluid diffusion with mechanical deformation, serve as excellent, i.e. "must-pass", benchmarks for verifying numerical models that aim at simulating fault behaviour under fluid injections.

In this study, we introduce an efficient and modular geomechanical solver for fluid injection with a focus on fault stability. The solver explicitly resolves both matrix and fractures. The algorithm follows the fully-implicit iterative approach proposed by \cite{prevost_partitioned_1997}, allowing the independent use of mechanics and fluid flow solvers. To our knowledge, this is the first time this approach is being used on a medium that explicitly contains discontinuities in the form of cracks having distinct permeability from the rock matrix. For simplicity, the fluid flow is assumed  to remain as single-phase. Both the mechanical and flow sub-problems are discretized spatially using finite elements, with the fracture mesh conforming to the matrix mesh. The accuracy of the solver is also verified using a novel analytical solution for the growth of fluid-induced fault slip.

The proposed geomechanical solver is built upon the open-source high-performance computing (HPC) finite-element library Akantu, developed at EPFL \citep{richart_akantu_2024,richart_implementation_2015,vocialta_3d_2017}. This software encompasses a broad spectrum of functionalities, including static and dynamic deformations of continuous bodies and structures, fracture growth and damage evolution, contact and friction, as well as heat and fluid flow. The HPC capabilities of Akantu are enabled by the use of the message passing interface (MPI) approach, facilitating efficient computational performance on complex simulations.

This paper is organized as follows. \cref{ch:theory} provides the equations governing deformation and fluid flow within porous media, faults, and fractures. The weak forms of these equations, along with details on the discretization of the matrix and fractures, complemented by the iterative solution scheme, are presented in \cref{ch:discretization}. \cref{ch:validation} discusses the evaluation of the accuracy and performance scalability of the solver through several benchmark cases. The practical application of the solver is showcased through a large-scale model of $\co2$ injection into a faulted aquifer, which is the focus of \cref{ch:application}. The appendix includes detailed instructions for the practical implementation of coupling operators required for the fully-implicit iterative solver.

\section{Theoretical model}
\label{ch:theory}
While the hydromechanical state of underground formations depends on multiple phenomena including deformation of the rock matrix, evolution of pressure and temperature, chemical interactions between the fluids and solids, in this article we focus on the rock deformations and pore pressure evolution as key factors. In the following subsections, we introduce the equations governing these phenomena both for the porous medium which comprises rock matrix and an enclosed fluid, as well as for the fractures and faults, which are treated differently due to their surface topology. We adopt a classical continuum approach, in which the pore-space saturated with a fluid and the rock matrix are treated as two overlapping continua.

\begin{figure}[htbp]
  \centering
  \includegraphics[width=0.5\columnwidth]{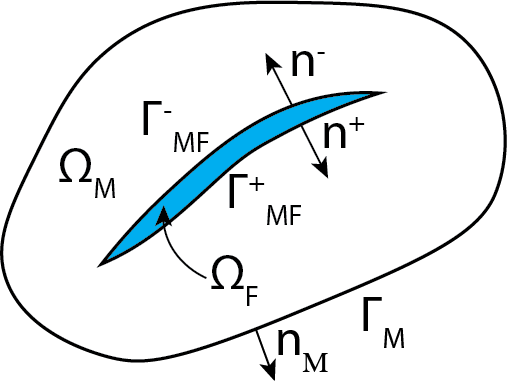}
  \caption{Physical domain split into the matrix $\Omega_M$ and fracture $\Omega_F$ volumes, separated by the fracture boundaries $\Gamma^{\pm}_{MF}$.}
  \label{fig:potato}
\end{figure}

We partition the domain $\Omega$ of the fractured porous medium into its continuous and discontinuous components. The continuum medium is identified as the matrix, denoted as $\Omega_M$, while the fracture volume is labelled as $\Omega_F$. The external boundary of the domain $\Omega$ is denoted as $\Gamma_{M}$, while the interface separating $\Omega_M$ and $\Omega_F$, representing the fracture surfaces, is designated as $\Gamma^{\pm}_{MF}$. The plus and minus superscripts are employed to distinguish the opposite fracture faces. The division of the physical domain is depicted in \cref{fig:potato}. Outward normals to the fracture surfaces are denoted by $n^{\pm}$.

\subsection{Porous medium}
In this study, we follow the formulation of poromechanics as described in \cite{coussy_poromechanics_2004,detournay_fundamentals_1993,rice_basic_1976} tracing back to the original work of \cite{biot_general_1941}. In this section, we use matrix notations. The governing equations for the porous medium and the pore fluid are respectively the balance of linear momentum and conservation of fluid volume. Under the assumptions of isotropic rock and quasistatic displacements, the governing equation for momentum balance of the porous rock is written as
\begin{equation} \label{eq:bal_mom}
    \nabla \cdot \boldsymbol{\sigma} + \rho_b \boldsymbol{g} = 0, \quad \mathbf{x} \in \Omega_M,
\end{equation}
where $\boldsymbol{\sigma}$ is the Cauchy total-stress tensor, $\boldsymbol{g}$ is the gravity vector, ${\rho_b = \rho_f \phi_0 + (1 - \phi_0)\rho_s}$ is the bulk density, $\phi_0$ is the initial porosity, and $\rho_s$ and $\rho_f$ are the solid and fluid densities, respectively. 

Following \cite{coussy_poromechanics_2004}, 
assuming small deformation, and linear poroelasticity from the reference state to the current state, 
the stress-strain poroelastic relation can be written as:
\begin{equation}\label{eq:stress}
     \stress - \stress_0 = \mathbb{C}: \boldsymbol{\varepsilon} - \biot (p - p_0) \boldsymbol{I} ,
\end{equation}
where $\mathbb{C}$ is the 4-th order stiffness tensor, $\boldsymbol{\varepsilon}$ is the strain tensor,  $\biot$ is the isotropic Biot coefficient, $p$ is the fluid pressure in the matrix, $\boldsymbol{I}$ is the identity tensor, and the subscript $0$ denotes the initial-state values. The linearized symmetric strain tensor is computed from the displacement field $\boldsymbol{u}$:
\begin{equation}\label{eq:strain}
    \boldsymbol{\varepsilon} = \frac{1}{2}(\nabla \boldsymbol{u} + \nabla^T \boldsymbol{u}).
\end{equation}


Failure of porous materials is driven by the Terzaghi effective stress $\stress'$ defined as:
\begin{equation}\label{eq:terzaghi}
 \stress =   \stress' -   p \boldsymbol{I}.
\end{equation}
%

The underground injection of $\co2$ is typically carried out into a reservoir containing brine, but it may also contain oil and gas. While $\co2$ is initially injected in a supercritical state, it may change phase to either liquid or gas within the reservoir, and it can also dissolve into brine. Modelling different phases and phase transitions explicitly requires the use of multiphase flow equations, which are well-documented in the literature \citep{aziz_petroleum_1979}. The formulation of single-phase fluid flow is a special case of the multiphase flow equations, which offers the advantage of reduced computational cost. A recent study by \cite{siriwardane_geomechanical_2016} demonstrated that the single-phase predictions of displacement and pressure in a $\co2$ injection scenario deviate by no more than $\qty{14}{\percent}$ from those of multiphase models. Given this, and in order to simplify the modelling process, we first opt for single-phase fluid flow which leads to the following equation of fluid volume conservation as \citep{detournay_fundamentals_1993}:
\begin{equation} \label{eq:con_mass}
  \frac{\partial \zeta}{\partial t} + \nabla \cdot \boldsymbol{q} = \gamma,   \quad \mathbf{x} \in \Omega_M,
\end{equation}
where $\zeta$ is the variation in the fluid content per unit of the saturated porous media. It is defined as the variation of fluid mass from the reference state divided by the fluid density in the reference state \citep{detournay_fundamentals_1993}. $\boldsymbol{q}$ is the fluid discharge vector (Darcy's velocity), and $\gamma$ is a possible volumetric source term. The variation of the fluid content depends on the volumetric deformation of the porous solid matrix as well as the fluid pressure change:
\begin{equation}\label{eq:fluid_content}
    \zeta = \biot \operatorname{tr} (\boldsymbol{\varepsilon}) + \frac{1}{M}(p - p_0),
\end{equation}
where $M$ is the Biot modulus, the reciprocal of which is the scaled sum of the intrinsic pore and fluid compressibilities as
\begin{equation}
    \frac{1}{M} = \frac{1}{N}  + \phi_0 c_f,
\end{equation}
where $N$ is the modulus linking the pressure variation and the porosity variation, $c_f$ is the fluid compressibility, and $\phi_0$ is the initial matrix porosity. For more details, see the seminal work of \cite{coussy_poromechanics_2004}. 

Darcy's law provide the expression of the fluid flux as function of the gradient of pore fluid pressure:
\begin{equation}\label{eq:darcy_matrix}
    \boldsymbol{q} = -\frac{k_M}{\mu_f} \nabla (p - p_0),
\end{equation}
where $k_M$ is the isotropic permeability of the rock matrix, $\mu_f$ is the fluid viscosity, and $p_0=\rho_f ||\boldsymbol{g}||$ denotes the hydrostatic pressure. In this study, we disregard variations in matrix permeability and fluid viscosity. Assuming the initial hydrostatic pressure distribution, we replace all the incremental values $(\stress - \stress_0)$
and $(p - p_0)$ by single terms $\stress$ and $p$, which now indicate the deviation of stress and pressure from the initial state.

In our numerical algorithm, we will discretize the flow equation in time by a backward difference scheme:
\begin{equation}
  \label{eq:time_disc}
        \frac{\zeta^{n+1} - \zeta^n}{\Delta t} + \nabla \cdot \boldsymbol{q}^{n+1} = \gamma^{n+1},
\end{equation}
where superscripts $(n+1)$ and $(n)$ denote the current and previous time steps, respectively, and $\Delta t$ is the duration of the time step.

\subsection{Fracture}
A fracture is defined as a discontinuity in the rock matrix delimited by two plane surfaces and filled with pore fluid. Given its planar topology, it requires different treatment for the definitions of deformation and pressure in the fracture. In this work, we will use the term "fracture" to denote pre-existing planar rock discontinuities occurring at various scales, including possibly large-scale faults.

\subsubsection{Fracture constitutive law} \label{ch:cohesive_law}
 The relative position of the fracture surfaces is defined by the opening vector $\boldsymbol{\delta} = \begin{Bmatrix}
     \open & \slip
 \end{Bmatrix}^T$, where $\open$ is the fracture normal opening and $\slip$ is the tangential displacement, or slip. The normal opening $\open$ has three possible configurations. The case $\open=0$ corresponds to a stress-free contact. When $\open>0$, two surfaces lose physical contact, and when $\open<0$, two surfaces are pressed into each other, which is termed "penetration". 

Similarly to Terzaghi effective stress in \cref{eq:terzaghi}, Terzaghi effective traction at the fracture surfaces $\boldsymbol{t}'$ is the sum of the total traction and fluid pressure acting along the fracture. Since $\boldsymbol{t}'$ is in balance with $\stress'$ within the matrix, it can be computed by projecting  $\stress'$ on the normal to the fracture surfaces $\boldsymbol{n}^{\pm}$ as:
\begin{equation}
    \boldsymbol{t}' = \boldsymbol{t} + p\boldsymbol{n}^{\pm} = - \boldsymbol{\sigma'} \boldsymbol{n}^{\pm},  \quad \mathbf{x} \in \Gamma_{MF}.
\end{equation}
The traction vector has its normal component $t'_n$ and the tangential one $t'_s$. Effective traction takes non-negative values only when fracture surfaces are in contact. The outward normals to the fracture surfaces $\boldsymbol{n}^{\pm}$ are defined in \cref{fig:potato}.

A constitutive law of a fracture or fault, representing an independent relationship between surface traction and fracture deformation, enables capturing fracture growth, opening, sliding and contact. Since the current study does not involve fracture growth, and the fracture surface is predetermined, the fracture geometry is defined in the beginning of the simulation. However, the employed finite-element code includes fracture growth functionality, allowing for a wide range of possibilities in future studies.

The normal component of the effective traction $t'_n$, which under compression becomes a contact pressure, is assumed to depend linearly on the normal opening:
\begin{equation}\label{eq:normal_traction}
    t'_n=k_n |\open| \quad \text{if \  $\open \leq 0$},
\end{equation}
where $k_n$ is the normal stiffness of contact. High amplitude of $k_n$ leads to low penetration values $\open$. Given that the rock was previously damaged during fracture formation, it is assumed that no tensile resistance exists, resulting in  $t'_n=0$ when $\open > 0$.
    
Further, we develop an elasto-plastic-like constitutive interfacial law to compute the shear fracture slip. To define the stability condition for a fault, we use the Mohr-Coulomb theory, symbolically expressed as:
\begin{equation}\label{eq:yield_function}
    \mathcal{F}(t') = |t'_s| - f t'_n   \leq 0 \quad \text{if \  $\open \leq 0$},
\end{equation}
where $\mathcal{F}$ is the yield function, and $f$ is the friction coefficient. When the fault is stuck, the magnitude of the shear traction is strictly below the shear strength $f t'_n$ and $\mathcal{F}<0$. Otherwise, if $\mathcal{F}=0$, shear traction reaches its maximum value and the frictional slip occurs. 

\begin{figure}[htbp]
  \centering
  \includegraphics[width=0.5\columnwidth]{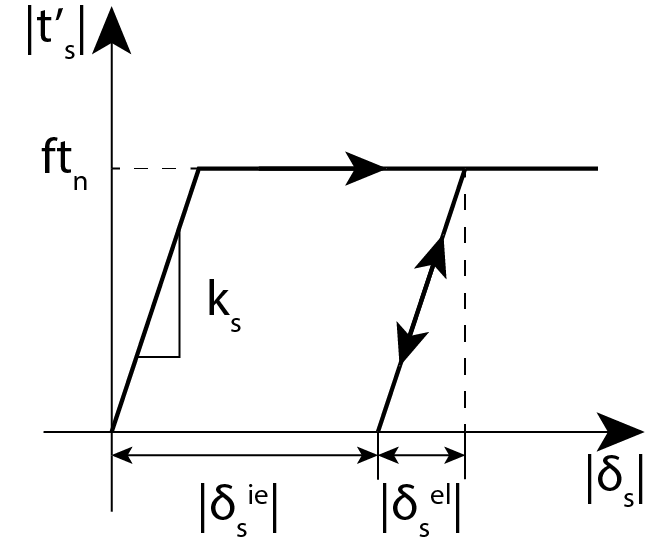}
  \caption{Slip-friction diagram for Mohr-Coulomb friction law}
  \label{fig:friction_law}
\end{figure}

This equation is classically numerically resolved with a regularization scheme employing a penalty coefficient $k_s$. Such regularization allows splitting the tangential slip into an elastic (stick or adhesive) part $\slip^{el}$ and inelastic (slip) part $\slip^{ie}$, as:
\begin{equation}\label{eq:slip_split}
    \slip = \slip^{el} + \slip^{ie}.
\end{equation}
\cref{fig:friction_law} depicts a typical friction-slip curve for the regularized elasto-plastic law defined by \cref{eq:yield_function}. If at a certain point, the fault is unloaded, the elastic slip will be recovered, while the inelastic slip will stay. The elastic part of the friction follows an isotropic linear elastic relation, which yields:
\begin{equation}
    t'_s = k_s \slip^{el},
\end{equation}
where $k_s$ is the elastic shear stiffness, which controls the rigidity of the fault and its capacity to deform elastically. Higher values of $k_s$ will reduce elastic deformation.

For a fixed contact pressure $t'_n$, a non-associated Mohr-Coulomb flow rule with no opening due to slip is utilized to characterize the evolution of the inelastic component of the slip:
\begin{equation}\label{eq:plastic_slip_increment}
    \Delta \boldsymbol{\slip}^{ie} = \Delta \lambda\frac{\partial \mathcal{G}}{\partial \boldsymbol{t}'_s} = \Delta \lambda \boldsymbol{\tau}, \quad \mathcal{G}(t) = ||\boldsymbol{t}'_s||,
\end{equation}
where $\boldsymbol{\tau}=\boldsymbol{t}'_s/||\boldsymbol{t}'_s||$ is the unit vector in direction of slip, $\mathcal{G}$ is the plastic flow potential, and $\Delta \lambda$ is the plastic multiplier increment. Furthermore, we incorporate the loading-unloading conditions in Karush-Kuhn-Tucker form as:
\begin{equation}\label{eq:kuhn-tucker}
    \Delta \lambda \geq 0, \quad \mathcal{F}(t') \leq 0, \quad \Delta \lambda \mathcal{F}(t') = 0,
\end{equation}
which dictate the increment of plastic parameter $\Delta \lambda$.

During each loading step, \cref{eq:yield_function,eq:slip_split,eq:plastic_slip_increment,eq:kuhn-tucker} are solved using the elastic-predictor plastic-corrector scheme. Initially, a trial traction $\boldsymbol{t}'^{,tr}$ is predicted under the assumption of purely elastic deformation. If this predicted traction fails to satisfy \cref{eq:yield_function} by overshooting the shear strength, it is plastically corrected by projecting the traction back onto the yield surface $\mathcal{F}$. In the case of Coulomb friction, this projection is done in a single step by computing directly $\Delta \lambda$ as:
\begin{equation}\label{eq:corrector}
    \Delta \lambda = \frac{1}{k_s}\left( |t'^{,tr}_s| - ft'_n\right).
\end{equation}
For further details on solving the frictional contact problem, readers may refer for example to the comprehensive textbook of \cite{wriggers_computational_2006}.

\subsubsection{Flow in the fracture}
Flow through the fracture is facilitated by its hydraulic aperture $w$, which can differ from the mechanical opening $\open$. Even in instances where the fracture is deemed closed, a microscopic conductive pathway persists, allowing fluid flow. Consequently, the hydraulic aperture $w$ aligns with the mechanical opening $\open$ for positive opening values, reverting to the default minimal value $w_0$, when the opening is zero or negative.

Within fractures, the same fluid volume conservation equation defined in \cref{eq:con_mass} applies, while the variation in the fluid content, due to the presence of the fluid phase only, changes to
\begin{equation} \label{eq:fluid_content_fracture}
  \zeta = \varepsilon^{\perp} + \frac{p}{M_F},  \quad \mathbf{x} \in \Omega_{F}.
\end{equation}
where $\varepsilon^{\perp}$ is the fracture deformation in its out-of-plane direction, $M_F$ is the Biot modulus within the fracture, and $p$ is the fluid pressure. The fluid flux in the fracture follows the same Darcy's law, but with anisotropic permeabilities. In this study, we distinguish between longitudinal permeability $k^{\parallel}$ and transverse permeability $k^{\perp}$, which enable the modelling of anisotropic fracture permeability as observed experimentally \citep{farrell_anisotropy_2014, zhang_effect_1998}. Therefore, Darcy's law for the fracture is written as:
\begin{equation}
    \label{eq:flux_decompose}
    \boldsymbol{q} = -\frac{k^{\parallel}}{\mu_f} \left( \frac{\partial p}{\partial x_1}\boldsymbol{n_1} + \frac{\partial p}{\partial x_2}\boldsymbol{n_2} \right) - \frac{k^{\perp}}{\mu_f} \frac{\partial p}{\partial x_3}\boldsymbol{n_3} ,
\end{equation}
where  $x_1$, $x_2$ represent two arbitrary axes laying on the fracture plane and perpendicular to each other, $x_3$ denotes the axis normal to the fracture, and $\boldsymbol{n_1}$, $\boldsymbol{n_2}$, and $\boldsymbol{n_3}$ are the corresponding unit vectors. After assuming linear variation of pressure within the fracture perpendicular to its plane, and denoting the pressure averaged along the $x_3$ axis as $\av{p}$ and the pressure jump along the same direction as $\jump{p}$, the equation above can be rewritten as
\begin{equation}
    \label{eq:flux_decompose_operators}
    \boldsymbol{q} = - \frac{k^{\parallel}}{\mu_f} \left(\frac{\partial \av{p}}{\partial x_1} \boldsymbol{n_1} + \frac{\partial \av{p}}{\partial x_2} \boldsymbol{n_2} \right) - \frac{k^{\perp}}{\mu_f} \frac{\jump{p}}{w} \boldsymbol{n_3} ,
\end{equation}
where $w$ is the hydraulic aperture of the fracture.

\section{Discretization and solution scheme}
\label{ch:discretization}
In this section, we discuss the numerical implementation of the geomechanical solver, covering discretization of the geometry, weak forms of the governing equations and the solution scheme.

\subsection{Discretization of the bulk and fractures}
Both solid and fluid subproblems are discretized using first-order finite elements. This discretization scheme was chosen for its simplicity; however, it is not restrictive, and different combinations of interpolation orders could be selected. For applications with high contrast in rock matrix stiffness and permeability, the inf-sup condition of Brezzi and Babuška \citep{brezzi_mixed_nodate, chapelle_inf-sup_1993} becomes particularly significant. To adhere to this condition, one should opt for second-order finite elements for the solid displacement and first-order for the pore-fluid pressure.

The proposed discretization scheme works both for the 2D and 3D problems, however, our focus here is on the 3D case. The matrix is discretized with tetrahedral elements, and the fractures are discretized with triangles. Alternatively, other standard elements could be utilized. The same discretized grid is employed for both the solid and fluid subproblems. 

In the solid part, the displacement variable $\disp$ is associated with the vertices of the elements, while the material properties such as Young modulus and Poisson ratio are assigned to the integration points. Similarly, in the fluid part, the pressure unknown $p$ is located at the vertices, while flow properties are defined at the integration points.

\subsubsection{Interface elements}
Fracture surfaces are discretized using interface elements, which are zero-thickness elements comprising two overlapping facets, resulting in a duplicated number of nodes. When elements of the same topology are used to dissipate energy upon opening, they are referred to as "cohesive" elements. Examples of their use can be found in the works of \cite{camacho_computational_1996, ortiz_finitedeformation_1999}.
\begin{figure}[htbp]
  \centering
  \includegraphics[width=0.3\columnwidth]{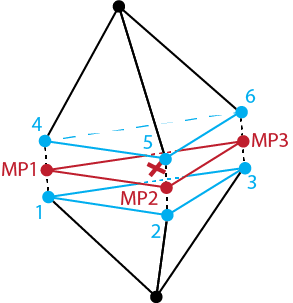}
  \caption{Interface element for linear tetrahedral elements}
  \label{fig:cohesive}
\end{figure}
These elements are inserted between the solid elements, necessitating compliance of the fracture surface with the solid mesh discretization. The schematic representation of such an element and its placement relative to solid elements is illustrated in \cref{fig:cohesive}, where duplicated nodes are highlighted in blue and the mid-plane, and fictitious mid-points of the interface element are shown in red. The normal to an element is calculated as an average between normals to two fracture facets. Traction vectors $\boldsymbol{t}$ are defined at the quadrature points of these interface elements, marked with a cross in \cref{fig:cohesive}. 

\subsubsection{Double-nodes flux approximation in the fracture}
To simulate flow within the fracture, we use the "double-nodes" approach introduced by \cite{segura_zero-thickness_2004}. This method involves employing interface elements with duplicated nodes to represent longitudinal flow along the fracture, transversal flow across the fracture, and the fluid flow between the matrix and the fracture in both directions.

As illustrated in \cref{fig:cohesive}, auxiliary nodes reside along the mid-plane of an interface element. At these nodes, the mean and the jump in pressure from \cref{eq:flux_decompose_operators} are defined and computed as
\begin{equation}\label{eq:operators}
\begin{split}
   &\av{p_{\textrm{MP}}} = \frac{1}{2}
   \begin{bmatrix} 
   I_p & I_p
   \end{bmatrix}
   p,\\
   &\jump{p_{\textrm{MP}}} = 
      \begin{bmatrix} 
   I_p & -I_p
   \end{bmatrix}
   p,
\end{split}
\end{equation}
where $I_p$ is an identity matrix with its order equal to the number of nodes in the interface element and $p$ is the vector of nodal pressures.

\subsection{Discretized forms of governing equations}
In this section, we establish the weak forms of governing equations, which are directly employed in the finite element method. The discretized forms and finite-element operators are expressed in a compact matrix form, as employed by \cite{logan_first_2012}, with matrices enclosed in square brackets $[\cdot]$, vectors in curly braces $\{\cdot\}$, and scalars without any brackets.

\subsubsection{Balance of momentum}
The weak form of the balance of momentum equation is obtained by pre-multiplying \cref{eq:bal_mom} with a vector test function $\{\testv\}$, substituting the constitutive relation \cref{eq:stress}, and integrating over the domain $\Omega_M$:
\begin{equation}
  \label{eq:discrete_balance}
  \begin{split}
    &\int_{\Omega_M} \{\eps(\testv)\}^T [C] \{\eps(\disp)\} \ d\Omega - \int_{\Omega_M} \{\eps(\testv)\}^T \biot p \{I\} \ d\Omega -
       \int_{\Gamma^+_{MF}\cup \Gamma^-_{MF}}  \av{\testv}^T  \{t^{\pm}_f\}  \ d\Gamma= \\
       &\int_{\Gamma_M} \{\testv\}^T \{t\}  \ d\Gamma_M  + \int_{\Omega_M}  \{\testv\}^T \rho_b \{g\}  \ d\Omega,
      \end{split}
\end{equation}
where $\{t^{\pm}_f\}$ are the total traction vectors acting along the opposing fracture surfaces $\Gamma^{\pm}_{MF}$ and having equal magnitudes but opposite directions, while $\{t\}$ represents the total traction acting on the external boundary of the domain $\Gamma_M$. The total traction on a fracture $\boldsymbol{t}_f$ results from the combined effects of the fluid pressure $p$ and the Terzaghi's effective stress $\boldsymbol{\sig}'$. For convenience, we express this equation in matrix notation as:
\begin{equation}
  \label{eq:total_traction}
  \boldsymbol{t}_f = - \left(\boldsymbol{\sig}' - \boldsymbol{I} p \right) \boldsymbol{n}^{\pm} = t'_n \boldsymbol{n}^{\pm} + t'_s \boldsymbol{\tau} + p \boldsymbol{n}^{\pm},
\end{equation}
%
where $t'_n$ and $t'_s$ are the effective normal and shear traction forces, $\boldsymbol{n}^{\pm}$ are the outward normals to the fracture surfaces, and $\boldsymbol{\tau}$ is the unit vector in the direction of slip. After substitution of \cref{eq:total_traction} into \cref{eq:discrete_balance}, we obtain the complete weak form of the equation of balance of momentum:
\begin{equation}
  \label{eq:discrete_balance_full}
  \begin{split}
    &\int_{\Omega_M} \{\eps(\testv)\}^T [C] \{\eps(\disp)\} \ d\Omega - \int_{\Omega_M} \{\eps(\testv)\}^T \biot p \{I\} \ d\Omega -
       \int_{\Gamma^+_{MF}\cup \Gamma^-_{MF}}  \av{\testv}^T (\{t'^{,\pm}_n\} + \{t'^{,\pm}_s\})  \ d\Gamma - \\
       &\int_{\Gamma^+_{MF}\cup \Gamma^-_{MF}}  \av{\testv}^T  \av{p} \{n^{\pm}\}  \ d\Gamma=
       \int_{\Gamma_M}   \{\testv\}^T  \{t\} \ d\Gamma  + \int_{\Omega_M}   \{\testv\}^T \rho_b \{g\} \ d\Omega.
      \end{split}
\end{equation}
In the above equation, both the normal $t'_n$ and shear traction $t'_s$ depend on the displacement field $\disp$ as described in \cref{ch:cohesive_law}.

\subsubsection{Fluid volume conservation in the matrix}
After applying a standard technique of multiplying with a scalar test function $r$, substituting the constitutive relations in \cref{eq:fluid_content,eq:darcy_matrix}, and performing integration by parts, we arrive to the weak form of the fluid volume conservation in the matrix:
\begin{equation}
  \label{eq:discrete_mass}
  \begin{split}
   & \int_{\Omega_M} r \biot \{I\}^T \frac{\partial \{\eps(\disp)\}}{\partial t} \ \text{d}\Omega + \frac{1}{M} \int_{\Omega_M} r\frac{\partial p}{\partial t}  \ \text{d}\Omega + 
    \int_{\Omega_M} \{\nabla r\}^T \frac{k_M}{\mu_f} \{\nabla p\} \ \text{d}\Omega = \\
   & -\int_{\Gamma_M} r \{q\}^T \{n_M\} \ \text{d}\Gamma - \int_{\Gamma^+_{MF} \cup \Gamma^-_{MF}} \av{r}(\{q^{+}\}^T \{n^+\} + \{q^{-}\}^T  \{n^-\}) \ \text{d}\Gamma + \int_{\Omega_M} r\gamma  \ \text{d}\Omega .
\end{split}
\end{equation}
In the above equation, $\{q\}$ is the incoming flux vector acting at the external boundary of the domain $\Gamma_M$ with its normal $\{n_M\}$, $\{q^{\pm}\}$ are the fluxes passing through the opposing fracture surfaces $\Gamma^+_{MF}$ and $\Gamma^-_{MF}$ with the corresponding normals $\{n^{\pm}\}$. Orientation of normals is illustrated in \cref{fig:potato}.

\subsubsection{Fluid volume conservation in the fracture}
By conducting the standard operation of plugging \cref{eq:flux_decompose_operators,eq:fluid_content_fracture} into the conservation of fluid volume equation \eqref{eq:con_mass}, multiplying it by a scalar test function $r$, integrating by parts, and averaging along the fracture width $w$, we obtain the following form:
\begin{equation}\label{eq:discrete_mass_fracture}
\begin{split}
    &\int_{\Gamma_{MF}} \av{r} \left(\frac{\partial w}{\partial t} + w S_F \frac{\partial \av{p}}{\partial t}\right) \text{d} \Gamma 
    +\int_{\Gamma_{MF}} \frac{w k^{\parallel}}{\mu} \left( \frac{\partial \av{r}}{\partial x_1}  \frac{\partial \av{p}}{\partial x_1} +  \frac{\partial \av{r}}{\partial x_2}  \frac{\partial \av{p}}{\partial x_2}\right) \ \text{d}\Gamma +\\
    & \int_{\Gamma_{MF}} \jump{r} \frac{k^{\perp}}{\mu w} \jump{p} \ \text{d} \Gamma = \int_{\Gamma^+_{MF} \cup \Gamma^-_{MF}} \av{r}(\{q^{+}\}^T \{n^+\} + \{q^{-}\}^T  \{n^-\}) \ \text{d}\Gamma + \\
    & \int_{\Gamma_{MF}} \av{r} \gamma w \ \text{d}\Gamma.
\end{split}
\end{equation}
%

When summing the weak forms of the fluid volume conservation equations \eqref{eq:discrete_mass} and \eqref{eq:discrete_mass_fracture}, the right-hand side terms of flux integrals over the opposing fracture surfaces cancel each other out.
The solution of the discretized \cref{eq:discrete_mass_fracture} is influenced by the fracture hydraulic transmissibility $wk^{\parallel}$. 

\subsection{Solution scheme}
\label{ch:bcg}
We solve the equations of balance of momentum and fluid volume conservation from a known solution at time $t_n$ to $t_n + \Delta t$ using a classical implicit integration. The unknowns are expressed in the incremental form as
\begin{equation}
    \begin{split}
        &\Delta u = u_{n+1} - u_n,\\
        &\Delta p = p_{n+1} - p_n.
    \end{split}
\end{equation}
We symbolically rewrite \cref{eq:discrete_balance_full,eq:discrete_mass,eq:discrete_mass_fracture} in the residual form as
\begin{equation}\label{eq:nonlinear_system}
    \begin{split}
    \boldsymbol{r}_m(\Delta u, \Delta p) &=    \boldsymbol{K}(u_{n+1}) \Delta u -\boldsymbol{A}_{\textrm{pu}}(u_{n+1}) \Delta p - f^u_{n+1} + \boldsymbol{K}(u_{n})u_n - \boldsymbol{A}_{\textrm{pu}}(u_{n}) p_n, \\
    \boldsymbol{r}_f(\Delta u, \Delta p) &= \boldsymbol{A}_{\textrm{up}}(u_{n+1}) \Delta u  + \left(\boldsymbol{S}(u_{n+1}) +\boldsymbol{C}(u_{n+1})\Delta t \right) \Delta p - \left(f^p_{n+1} - \boldsymbol{C}(u_{n}) p_n \right) \Delta t,
    \end{split}
\end{equation}
where $\boldsymbol{r}_m$ and $\boldsymbol{r}_f$ are the residuals of the mechanical and fluid flow problems correspondingly, $\boldsymbol{K}$, $\boldsymbol{S}$, $\boldsymbol{C}$ are the tangent stiffness, storage and conductivity matrices, $\boldsymbol{A}_{\textrm{pu}}$ is the pressure-to-displacement coupling matrix, $\boldsymbol{A}_{\textrm{up}}$ is the displacement-to-pressure coupling matrix, $f^u$ is the vector of nodal loads accounting for the gravity force and Neumann boundary condition, and $f^p$ is the vector of flow sources comprising the incoming fluxes over the external boundary and potential source/sink terms.

The resulting system \eqref{eq:nonlinear_system} is a non-linear system of equations due to the inclusion of frictional slip, which is reflected in the operator $\boldsymbol{K}$, as well as the variation in orientations of the opposing fracture surfaces $\Gamma^+_{MF}$ and $\Gamma^-_{MF}$ reflected in their respective normal vectors $\boldsymbol{n}_n^{\pm}$. The normals directly affect integrals of pressure $p$ and traction $\boldsymbol{t}$ along the fracture surfaces in \cref{eq:discrete_balance_full}, as well as the computation of fracture deformation speed $\partial w / \partial t$ in \cref{eq:discrete_mass_fracture}, which is transmitted through the operators $\boldsymbol{A}_{\textrm{pu}}$ and $\boldsymbol{A}_{\textrm{up}}$, rendering them non-linear. In a more general case, storage and permeability of the rock matrix and fractures could also vary with deformation and displacement discontinuity across the fractures, rendering the operators $\boldsymbol{S}$ and $\boldsymbol{C}$ non-linear. However, for the purposes of this study, we assume that they remain constant.

\subsubsection{Iterative solution of the non-linear system}
We use a Newton-Raphson scheme to find the root of such a non-linear system of equations combined with an elastic predictor-corrector step to integrate locally the interface elasto-plastic law. In each iteration, the method linearizes the system around the current guess, solves the linearized system for a correction, and updates the guess. This iterative update of the solution can be expressed as
\begin{equation}
    \begin{split}
        &\Delta u^{i+1} = \Delta u^{i} + \text{d} u^i,\\
        &\Delta p^{i+1} = \Delta p^{i} + \text{d} p^i,
    \end{split}
\end{equation}
where $\Delta u^{i+1}$ and $\Delta p^{i+1}$ are the solution vectors to the system updated after the $i$-th Newton iteration, and $\text{d}u^i$ and $\text{d}p^i$ are the last corrections. These corrections are computed by solving the following coupled Jacobian linear system of equations:
\begin{equation}\label{eq:linear_system}
    \begin{bmatrix}
    \boldsymbol{K} & - \boldsymbol{A}_{\textrm{pu}} \\
    \boldsymbol{A}_{\textrm{up}} & \boldsymbol{S}+\boldsymbol{C}\Delta t
    \end{bmatrix}
    \begin{Bmatrix}
    \text{d}u^i_{n+1}\\
    \text{d}p^i_{n+1}
    \end{Bmatrix}
    =
    \begin{Bmatrix}
    &f^u_{n+1} - \boldsymbol{K}u_{n+1}^i + \boldsymbol{A}_{\textrm{pu}} p_{n+1}^i \\
    &\left(f^p_{n+1} - \boldsymbol{C} p_{n+1}^i \right)\Delta t - \boldsymbol{S} \Delta p_{n+1}^i - \boldsymbol{A}_{\textrm{up}} \Delta u_{n+1}^i
    \end{Bmatrix},
\end{equation}
where the matrix on the left is the block-structured Jacobian matrix.

\subsubsection{Iterative solution of the Jacobian system}
Using the full matrix system while solving the set of linear equations \eqref{eq:linear_system} is computationally burdensome for three-dimensional problems, necessitating extensive assembly and factorization of the Jacobian matrix. Furthermore, a specialized solver integrating two field equations has to be developed, hindering direct utilization of separate solvers specialized in pressure diffusion and static deformations. To overcome this obstacle, we solve the linear system by a partitioned iterative conjugate gradient (CG) approach proposed by \cite{prevost_partitioned_1997}, who claimed it to be unconditionally stable, robust and computationally efficient for the simultaneous integration of transient coupled field problems. 

Adopting the Prevost's notation, we replace \cref{eq:linear_system} by
\begin{equation}\label{eq:prevost_system}
    \begin{bmatrix}
    \boldsymbol{A}_{11} & \boldsymbol{A}_{12} \\
    \boldsymbol{A}_{21} & \boldsymbol{A}_{22}
    \end{bmatrix}
    \begin{Bmatrix}
    \boldsymbol{x}_1\\
    \boldsymbol{x}_2
    \end{Bmatrix}
    =
    \begin{Bmatrix}
    \boldsymbol{b}_1 \\
    \boldsymbol{b}_2
    \end{Bmatrix},
\end{equation}
where $\boldsymbol{A}_{11} = \boldsymbol{K}$, $\boldsymbol{A}_{22} = \boldsymbol{S} + \boldsymbol{C}\Delta t$, $\boldsymbol{A}_{12} = -\boldsymbol{A}_{\textrm{pu}}$, $\boldsymbol{A}_{21} = \boldsymbol{A}_{\textrm{up}} $, $\boldsymbol{x}_1$ and $\boldsymbol{x}_2$ are two respective unknowns, and $\boldsymbol{b}_1$ and $\boldsymbol{b}_2$ are the residuals. A direct partitioned solution of \cref{eq:prevost_system} requires assembly of the Schur complement matrix. This is done by deriving $\boldsymbol{x}_1$ from the first line of \cref{eq:prevost_system} as
\begin{equation}
    \boldsymbol{x}_1 = \boldsymbol{A}_{11}^{-1} (\boldsymbol{b}_1 - \boldsymbol{A}_{12} \boldsymbol{x}_2)
\end{equation}
and later substituting it into the second equation, which results in
\begin{equation} \label{eq:schur_system}
    (\boldsymbol{A}_{22} - \boldsymbol{A}_{21} \boldsymbol{A}_{11}^{-1} \boldsymbol{A}_{12})\boldsymbol{x}_2 = \tilde{\boldsymbol{S}}\boldsymbol{x}_2 = \boldsymbol{b}_2 -\boldsymbol{A}_{21} \boldsymbol{A}_{11}^{-1} \boldsymbol{b}_1,
\end{equation}
where $\tilde{\boldsymbol{S}}$ is the symmetric-positive Schur complement. Prevost, in his method, never builds $\tilde{\boldsymbol{S}}$ explicitly and solves \cref{eq:schur_system} by the symmetric CG method \citep{hestenes_methods_1952} preconditioned with $\boldsymbol{A}_{22}^{-1}$. Therefore, the proposed procedure consists of iteratively solving subproblems with respective Jacobian matrices $\boldsymbol{A}_{11}$ and $\boldsymbol{A}_{22}$, which is handled by separate fluid flow and deformation solvers. A comprehensive review of the Prevost's method and a comparison of its efficiency with that of direct and projected solvers for coupled systems of equations is detailed in \cite{gambolati_direct_2002}.

Matrices $\boldsymbol{A}_{11}$ and $\boldsymbol{A}_{22}$ are assembled by their respective single-field solvers. The coupling matrices $\boldsymbol{A}_{12}$ and $\boldsymbol{A}_{21}$ are never assembled, and their matrix-vector products of the $\boldsymbol{A}\boldsymbol{p}$-type are directly evaluated at the level of finite elements and later assembled into the global vectors. While examples in the literature demonstrate the usage of Prevost's method for addressing issues within continuous fields \citep{gambolati_direct_2002,kalateh_finite_2018}, to the best of our knowledge, this study marks the first application to a problem incorporating displacement and pressure field discontinuity in a form of fractures. The main challenge lies in  implementing the $\boldsymbol{A}\boldsymbol{p}$ products for the fracture elements, which are two-dimensional structures embedded between three-dimensional solid elements. Details on these implementations in the context of finite elements are given in \ref{app:coupling} and \ref{app:coupling_fracture}.

\section{Model verification}
\label{ch:validation}
\subsection{Cryer's sphere problem}
In this first test, we verify the model's ability to reproduce the Mandel-Cryer effect \citep{cryer_comparison_1963,mandel_consolidation_1953}. This phenomenon involves a non-monotonous evolution of pore-pressure under a constant mechanical load. Such an effect is intrinsically linked to a time-dependent transfer of porelastic strain, and as such can be reproduced only by accounting for the full poroelastic coupling. 

\begin{figure}[]
\centering
  \begin{subfigure}{0.4\textwidth}
    \includegraphics[width=\columnwidth]{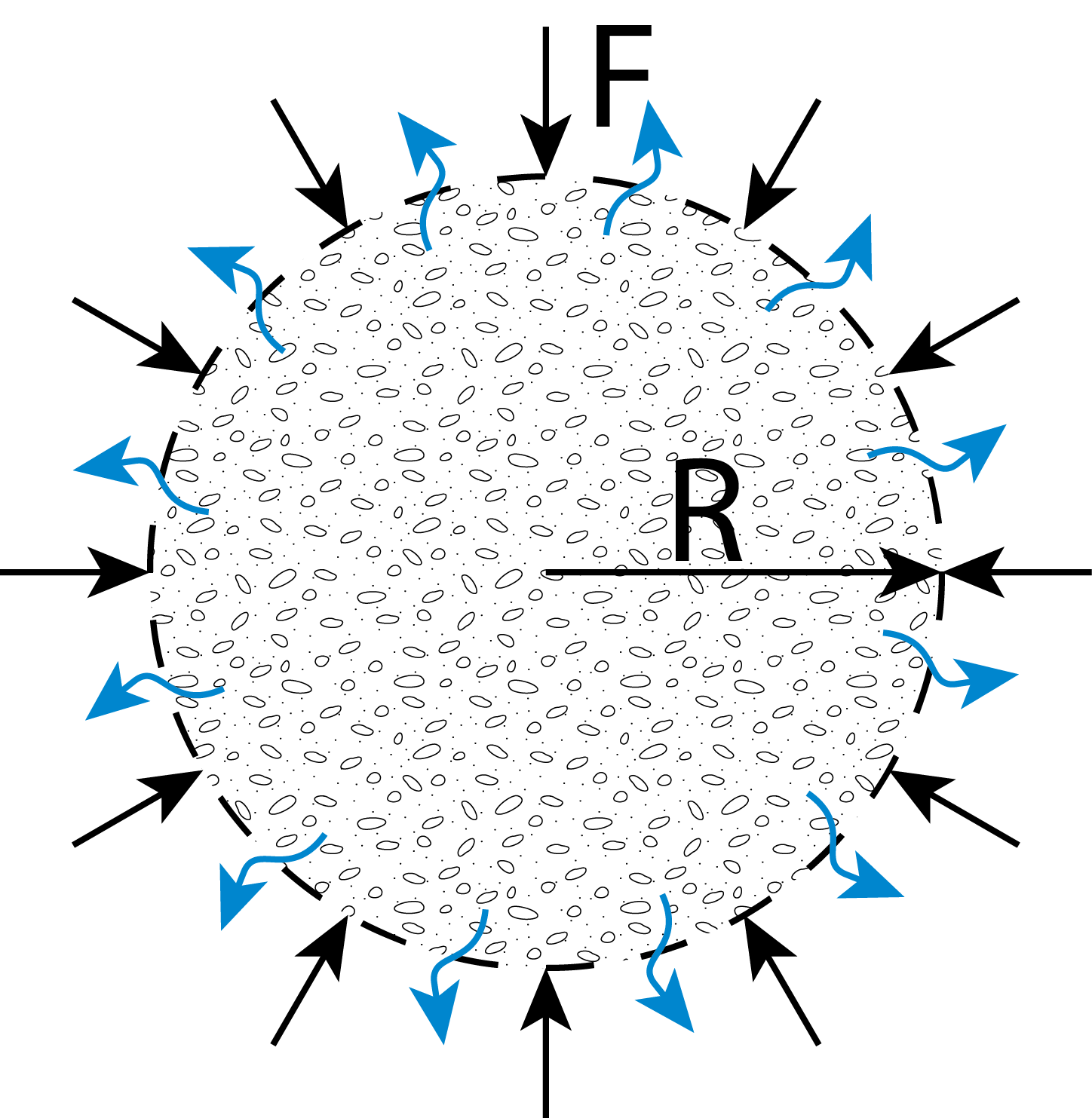}
  \end{subfigure}
  \begin{subfigure}{0.5\textwidth}
    \includegraphics[width=\columnwidth]{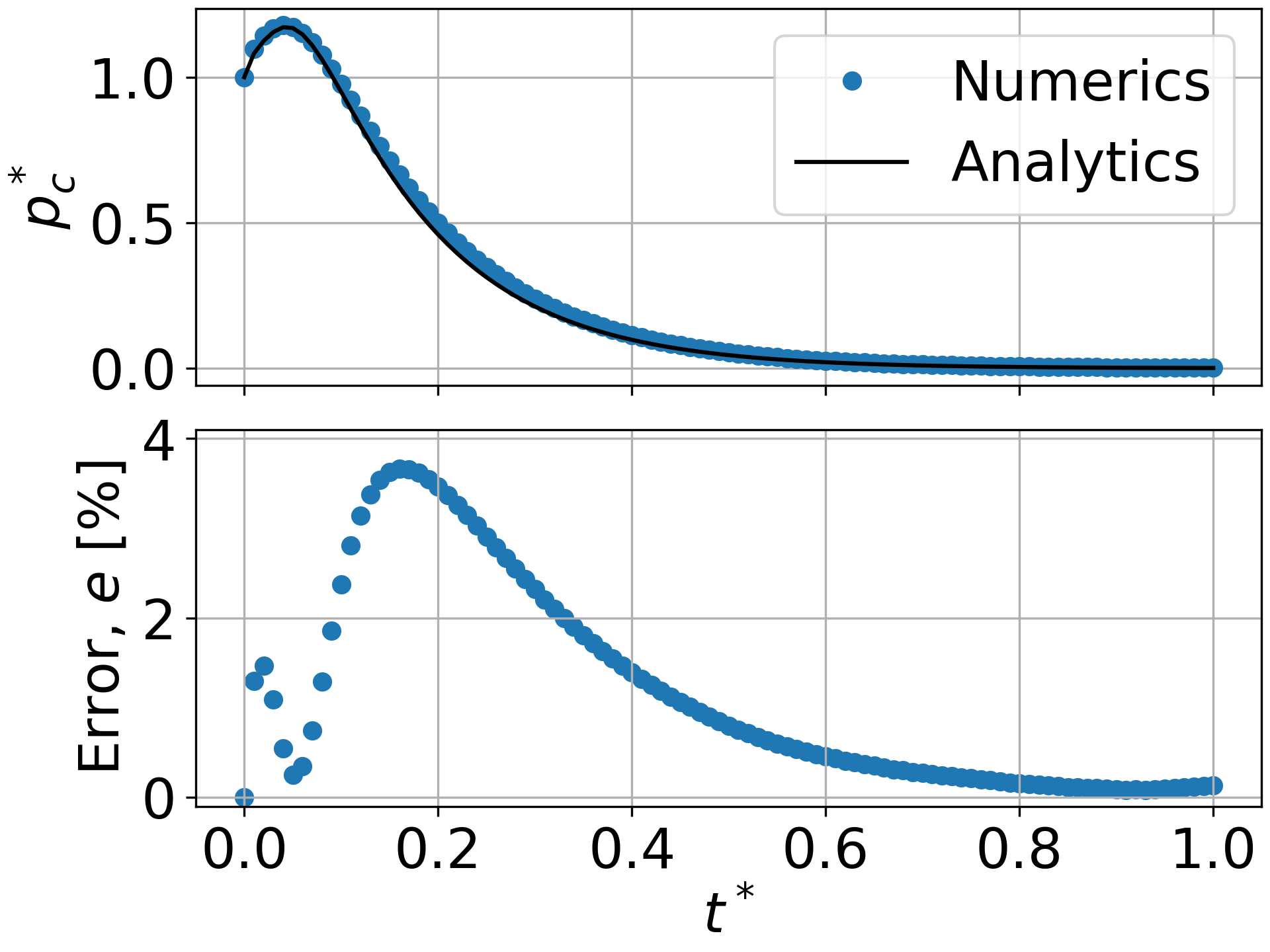}
  \end{subfigure}
    \caption{Setup of the Cryer's sphere problem with boundary conditions (\textit{left}). Evolution of the pore pressure at the centre of the sphere (analytical, numerical results, and the absolute error scaled with respect to the peak pressure) (\textit{right}). 
  }
  \label{fig:cryer}
\end{figure}

The Cryer's problem involves a homogeneous permeable poroelastic sphere of radius $R$, saturated with fluid initially at zero pressure, as depicted in \cref{fig:cryer}. The fluid can freely drain from the outer surface of the sphere. When subjected to a uniform radial compressive loading $F$ applied at its surface, the sphere undergoes  instantaneous compression. The resulting undrained response leads to a uniform pressure build-up of the amplitude $p_0$ throughout the sphere. Subsequently, fluid pressure begins to drain from the outer surface, causing softening in the outer layer of the sphere. This decrease in stiffness leads to stress redistribution across the sphere, leading to further compression of the inner region still experiencing undrained conditions. At the centre of the sphere, two opposing effects occur: pressure diffusion and pressure build-up due to continuous compression of the inner part. This phenomenon is illustrated in the right part of \cref{fig:cryer}, where the dimensionless pressure at the sphere's centre $p^*_c = p(r=0,t)/p_0$ is plotted against the dimensionless time $t^*=c_v t /R^2$, where $c_v = {k_M M}/{\mu_f}$ represents the hydraulic diffusivity, and $M$ is the Biot modulus. Initially, pressure build-up prevails, but at later stages, diffusion dominates, causing pressure to dissipate towards zero. A closed form analytical solution to this problem is available in \cite{verruijt_theory_2010,cheng_poroelasticity_2016}.

The numerical model of the Cryer's problem only contains one quarter of the sphere thanks to the problem's symmetry. The current 3D geometry, illustrated in \cref{fig:scalability}, contains $393,340$ nodes, which corresponds to $1,573,360$ DOFs comprising both pressure and displacement. The error estimate, depicted in the right section of \cref{fig:cryer}, comprises the absolute error normalized with respect to the peak pressure as $e = (p^{*\text{num}}_c - p^{*\text{an}}_c)/p^{*\text{an,max}}_c$. The numerical solution accurately predicts the evolution of the pore-fluid pressure, with the relative error residing below $4\%$. A closer match with the analytical solution can of course be achieved by refining the mesh and the time-step size.

\subsubsection{Computational scalability}
\begin{figure}[]
\centering
    \begin{subfigure}{0.4\textwidth}
        \includegraphics[width=\columnwidth]{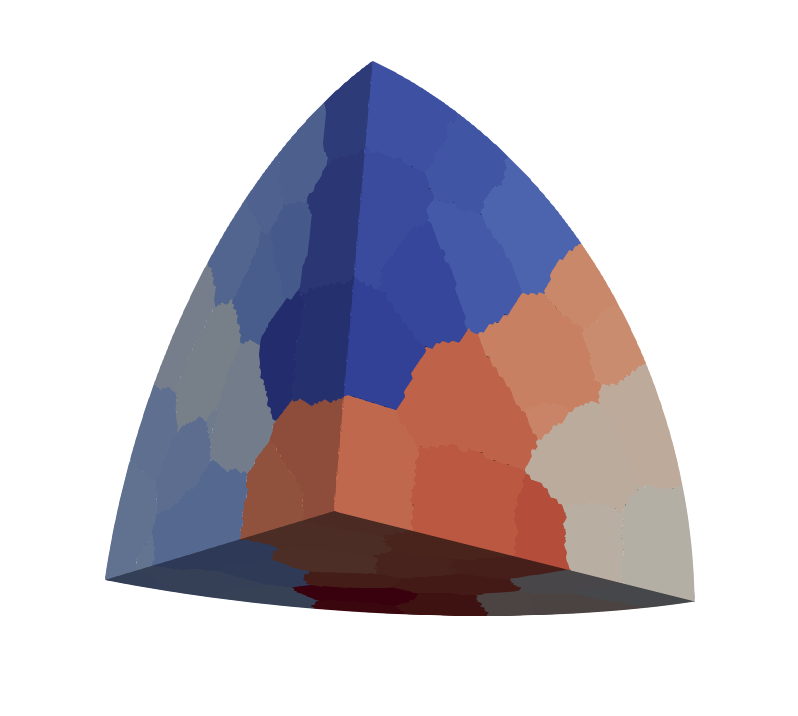}
    \end{subfigure}
    \begin{subfigure}{0.5\textwidth}
        \includegraphics[width=\columnwidth]{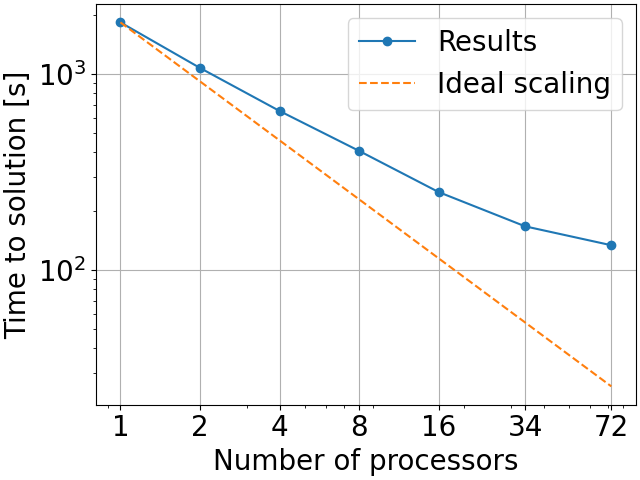}
    \end{subfigure}
    \caption{Partitioning of the Cryer's sphere mesh with $1,573,360$ degrees of freedom across $72$ processors in the scalability study (\textit{left}). Strong scaling of the solver compared to the ideal linear scaling (\textit{right}). As the mesh used is relatively coarse, the gain in time to solution with the number of process asymptote at about 72 processors. A finer mesh will shift the beginning of sublinear scaling to a larger number of processors accordingly.}
    \label{fig:scalability}
\end{figure}

We also use the  Cryer's problem to study the scalability of the developed parallel solver. We discuss here only the strong scaling which evaluates how the simulation time changes with the increase in the number of processors used for a problem with a fixed number of degrees of freedom (DOFs). The same simulation was run on 1 to 72 processors, on a cluster of Intel Xeon nodes with two sockets of 36 processors, 512Gb of RAM and dual 25Gb Ethernet links. The mesh and its partitioning across $72$ processors are depicted in the left part of \cref{fig:scalability}. A single time step with $40$ CG iterations was timed, and the time to solution for each number of processors is illustrated in the right section of \cref{fig:scalability}. The orange line in the plot represents the ideal scaling. The solver exhibits sublinear scaling, starting with an exponent of $0.85$ up to $16$ processors, beyond which the scalability gradually decreases. This trend is attributed to the extra communications involved in computing the coupling terms of $\boldsymbol{Ap}$-type within the iterative partitioned CG solution described in \cref{ch:bcg}. For a larger mesh with smaller elements, this reduction in scalability will happen for a larger number of processors.

\subsection{Injection-induced fault slip growth}
\label{ch:sheared_fault}
The second verification test assesses the model's capability to accurately predict shear slip activation of a fault due to fluid injection into it. This scenario is particularly important in geothermal operations in relation to hydraulic stimulation of pre-existing fractures to enhance the permeability of the rock mass. It can also occur in the context of waste-water injection in reservoir as well as $\co2$ storage. 

\begin{figure}[H]
\centering
  \begin{subfigure}{0.7\textwidth}
    \includegraphics[width=\columnwidth]{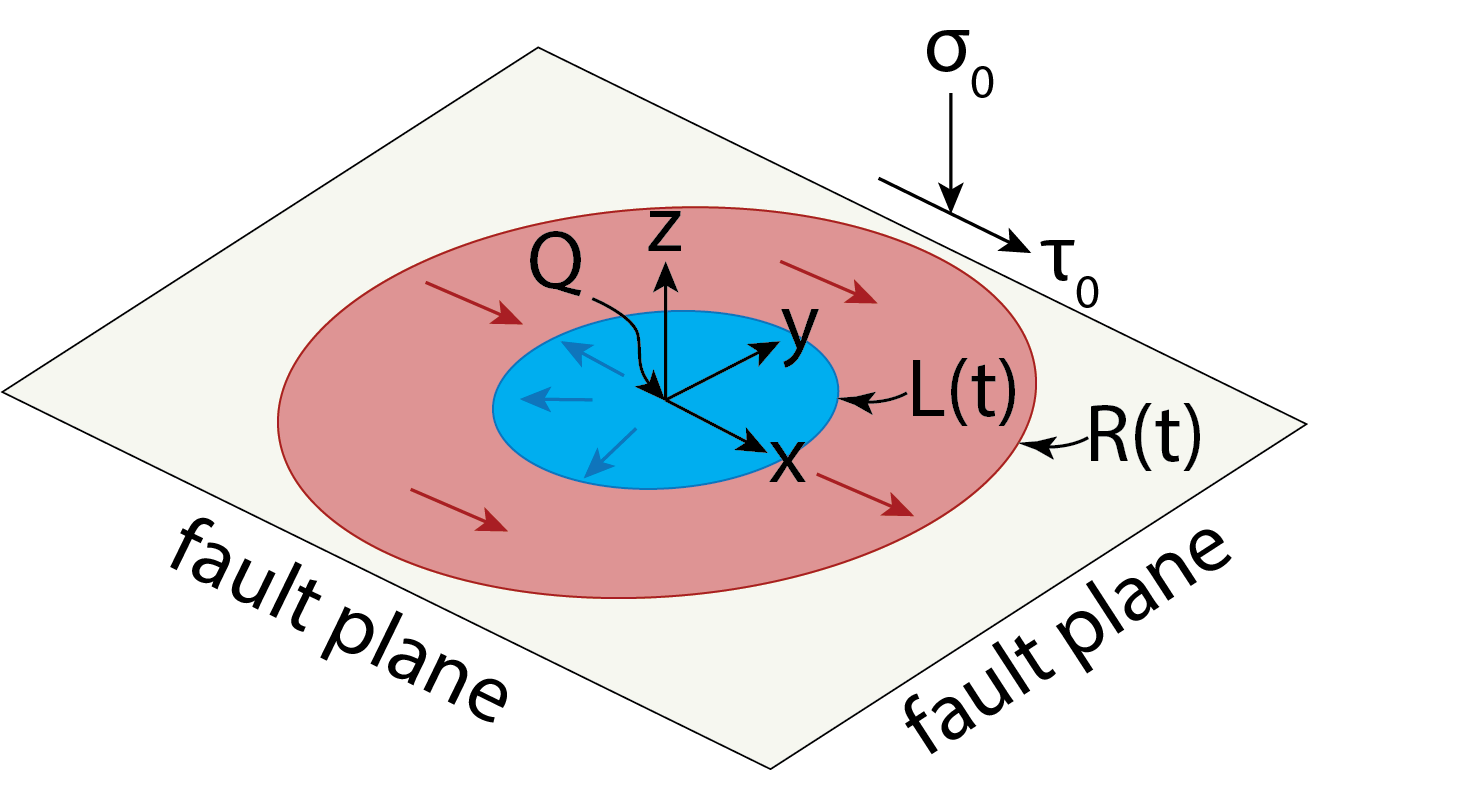}
  \end{subfigure}
  \begin{subfigure}{0.7\textwidth}
    \includegraphics[width=\columnwidth]{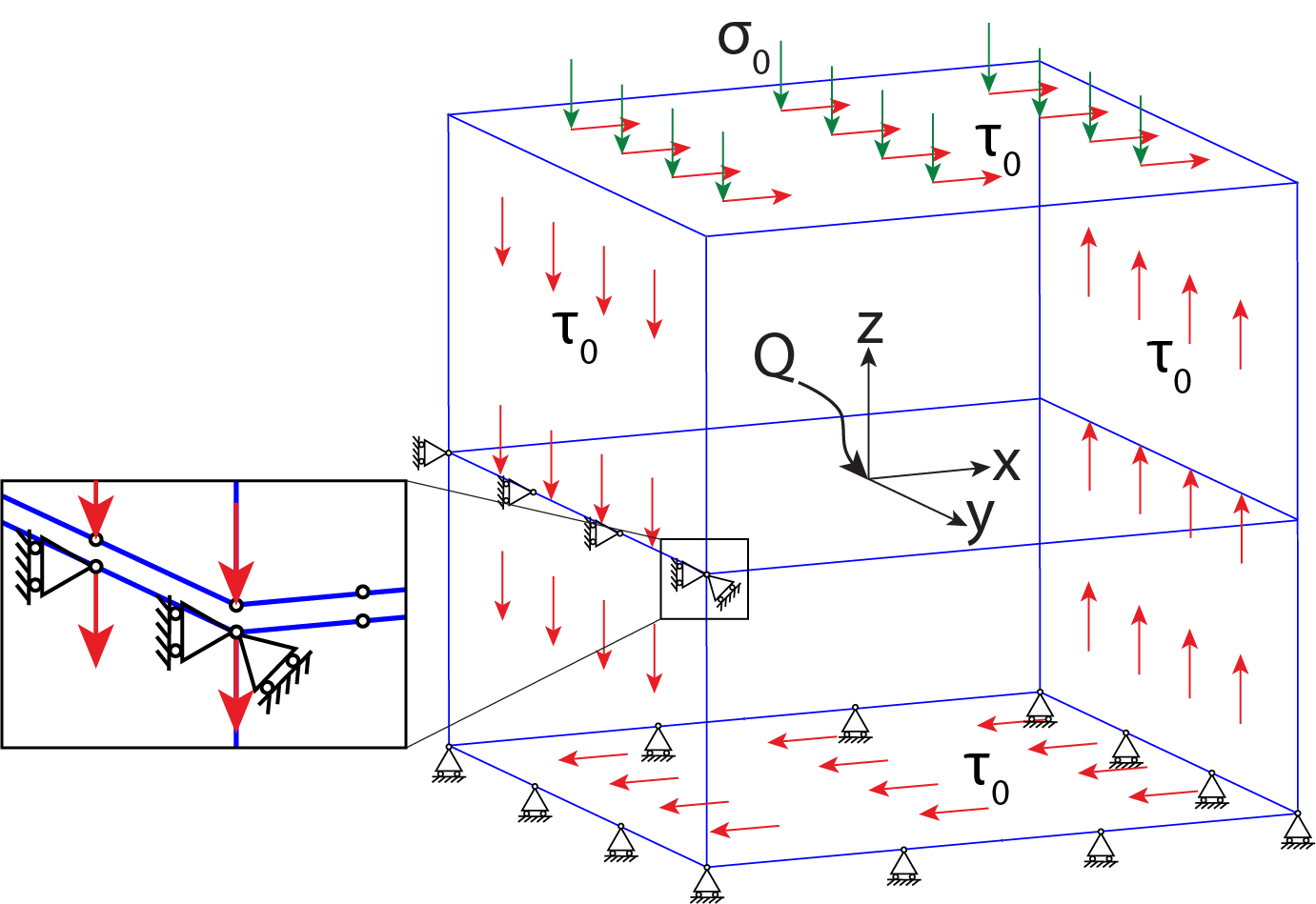}
  \end{subfigure}
  \caption{Configuration of the injection into a pre-existing frictional fracture, adapted from \cite{saez_three-dimensional_2022} (\textit{top}). Numerical setup of the injection into a fracture (\textit{bottom}). Due to the insertion of interface elements, the nodes along the whole fracture plane are duplicated. The artificial opening shown in the zoomed-in section is included for visualisation purposes \textit{only}. To achieve a homogeneous shear stress along the fracture, one row of nodes on one side of the fracture is restrained in the $x$ direction (see inset).}
  \label{fig:fault}
\end{figure}

The verification configuration comprises an infinite planar shear fracture confined by two semi-infinite homogeneous isotropic and linear elastic solids, as depicted in the top section of \cref{fig:fault}. Initially, the fracture homogeneously experiences a normal stress $\sigma_0$ along the $z$-axis and a shear stress $\tau_0$ along the $x$-axis. The behaviour of the fracture follows Coulomb's friction law with a constant friction coefficient. Fluid is continuously injected at a constant rate $Q$ at the origin which unclamps the fault by lowering the effective normal stress. Assuming an impermeable solid matrix, a uniform and constant permeability of the fracture, fluid flow remains confined to the fracture surface, and the pressure profile has a peak value at the origin and axisymmetric distribution with respect to the $z$-axis. The radius of the circular pore-pressure front, denoted as $L(t)$, and the pressure profile, have well-established analytical solutions \citep{theis_relation_1935,cheng_poroelasticity_2016}. The increase in pore pressure induces fracture slip $\delta$, with its maximum value occurring at the origin and gradually diminishing towards the rupture front $R(t)$. 

An analytical solution for both the rupture front and the slip profile was obtained for a 2D plane-strain configuration by \cite{bhattacharya_fluid-induced_2019,viesca_self-similar_2021}. 
The more realistic case of an injection into a three-dimensional planar fault was investigated in details
by \cite{saez_three-dimensional_2022}, who notably obtained an analytical solution for the case of a circular rupture (strictly valid for the case of a zero Poisson's ratio). 
\cite{saez_three-dimensional_2022} have notably shown that the rupture grows in a self-similar manner. The rupture radius $R(t)$ expands proportionally to the pore-pressure diffusion front $L(t)$: $R(t)=\lambda L(t)$. The amplification factor $\lambda=R(t)/L(t)$ depends on a dimensionless fault-stress injection parameter $T$. This parameter is defined as the ratio between the distance to failure under ambient conditions and the strength of the injection:
\begin{equation}
    T = \frac{1 - \tau_0/f\sigma'_0}{\Delta p_*/\sigma'_0},
\end{equation}
where $\tau_0$ is the initial shear stress along the fault, $\sigma'_0$ the initial normal effective stress (taken positive in compression), $\Delta p_*= Q \mu_f / (4 \pi k^{||} w)$ is the characteristic fluid pressure induced by the injection at a constant volumetric rate $Q$, and $f$ is the fault friction coefficient. 
When the fault injection stress parameter is small ($T \ll 1$), the fault is said to be critically stressed (CS) as either the initial distance to failure is small or the injection strength is large. In that case, 
the rupture front $R(t)$ is significantly ahead of the diffusion front $L(t)$ ($\lambda \gg 1$). 
Alternatively, when the fault injection stress parameter is large ($T \gg 1$ in relation to either a large initial distance to failure or a weak injection strength), the rupture front lies within the pressurized region ($\lambda \ll 1$) such that this limit is denoted as marginally pressurized (MP). 
Closed form analytical expressions for $\lambda$ and $\delta(r,t)$ were derived for the limiting cases that are critically stressed (CS) with $T \ll 1$ and marginally pressurized (MP) faults with $T \gg 1$ (\cite{saez_three-dimensional_2022,viesca_asymptotic_2024} for details).

In the current study, we verify the accuracy of the developed solver for both limiting cases against these recently obtained analytical solutions for fluid-induced frictional rupture growth. The geometry of the numerical model contains a solid rectangular prism with dimensions $\qtyproduct{40 x 40 x 40}{\meter}$, which is depicted in the bottom part of \cref{fig:fault}.
A horizontal fracture is passing through the middle plane of the prism at $z = 0$. The initial stress state along the interface was established by applying compressive normal and shear tractions on the four prism boundaries. For the boundary conditions, we have restrained the vertical displacement of the bottom surface, as well as the horizontal displacement of the lower half of duplicated cohesive nodes on the left side of the box. Injection $Q$ was applied at the origin at a pair of cohesive nodes by splitting the injection rate equally between them. The meshing of the fracture is irregular to avoid any directional bias, and the average size of a interface element is $\qty{0.05}{\meter}$. Material properties of the solid medium, fluid and fracture are given in \cref{tab:material_prop}.
\begin{table}[H]
  \centering
  \begin{tabular}{l r}
    \toprule
    Young modulus $E$, $\unit{\giga \pascal}$ & $60$\\
    Poisson ratio $\nu$, - & $0$\\
    Matrix permeability $k_M$, mD  & $0$\\
    Biot coefficient $\alpha$, - & $0$\\
    Biot modulus $M$, $\unit{\giga \pascal}$ & \num{1e5}\\
    Fluid viscosity $\mu_f$, $\unit{\milli \pascal \second}$ & \num{0.89}\\
    Fracture friction coefficient $f$, - & $\num{0.6}$\\
    Fracture normal stiffness $k_n$, $\unit{\giga \pascal}$ & \num{6e3}\\
    Fracture shear stiffness $k_s$, \unit{\giga \pascal} & \num{6e2}\\
    Fracture hydraulic transmissibility $wk^{\parallel}$, \unit{\cubic \meter} & \num{1e-15}\\
    Fracture perpendicular permeability $k^{\perp}$, mD & $\infty$ \\
    Fracture Biot modulus $M_F$, \unit{\mega \pascal} & \num{0.2}\\ 
    \toprule
  \end{tabular}
  \caption{Material properties of the rock, fluid, and fracture}
  \label{tab:material_prop}
\end{table}

The two simulated scenarios correspond to the limiting cases of the analytical solution. The critically stressed case, characterized by $T=0.01$, is modelled using the following loading parameters: a vertical load $\sigma_0= \qty{80}{\mega \pascal}$, a shear load $\tau_0=\qty{47.999}{\mega \pascal}$, and a final time $t_f=\qty{120}{\second}$. The marginally pressurized case, with $T=4$, is established by slightly reducing the shear load to $\tau_0=\qty{47.830}{\mega \pascal}$, and increasing the final time to $t_f=\qty{5400}{\second}$. In both cases, the injection rate remained constant at $Q=\qty{0.06}{\liter \per \minute}$.

\begin{figure}
\begin{center}
  \begin{subfigure}{0.47\textwidth}
    \includegraphics[width=\columnwidth]{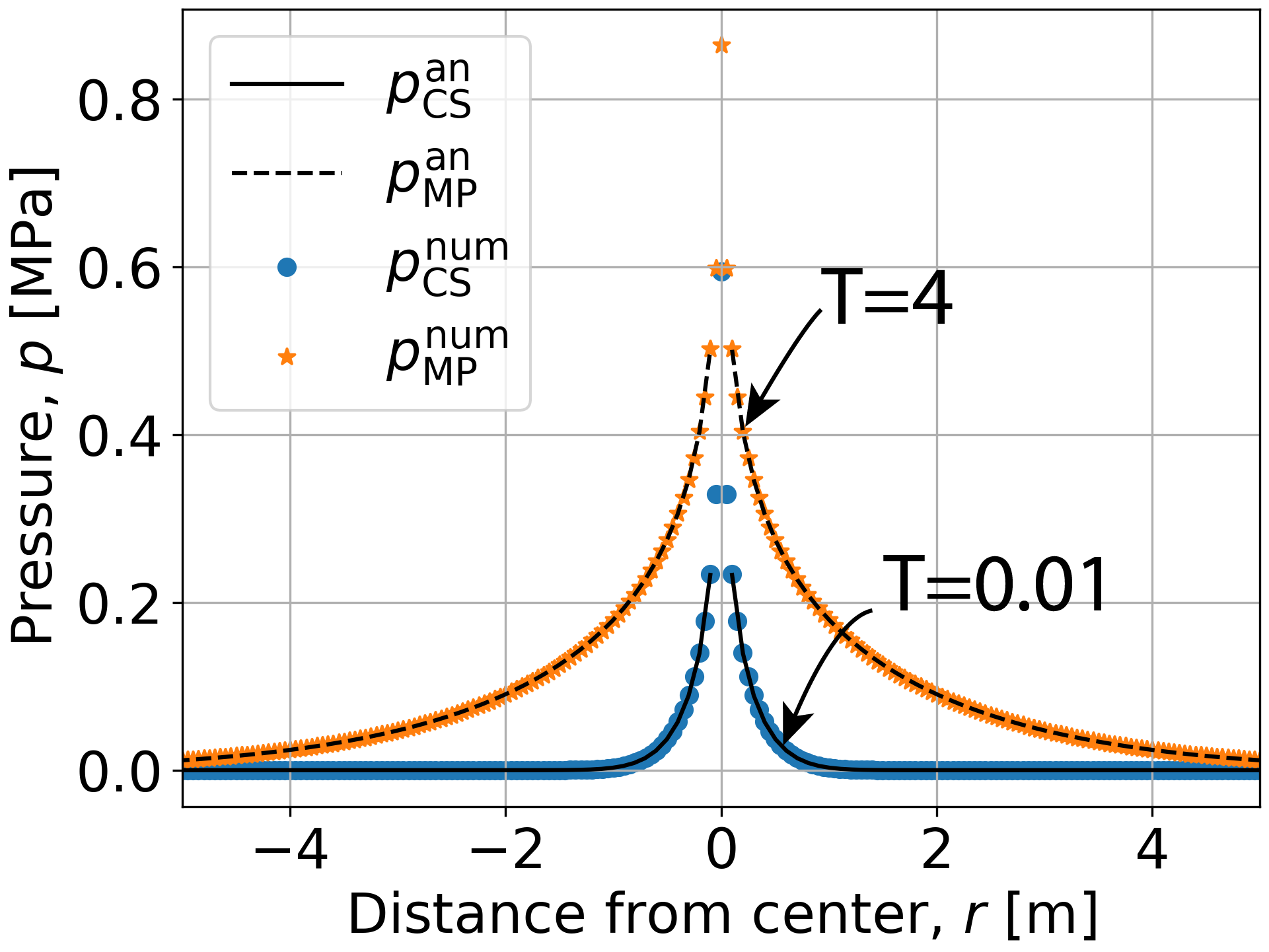}
  \end{subfigure} \hspace{0.2cm}
  \begin{subfigure}{0.49\textwidth}
    \includegraphics[width=\columnwidth]{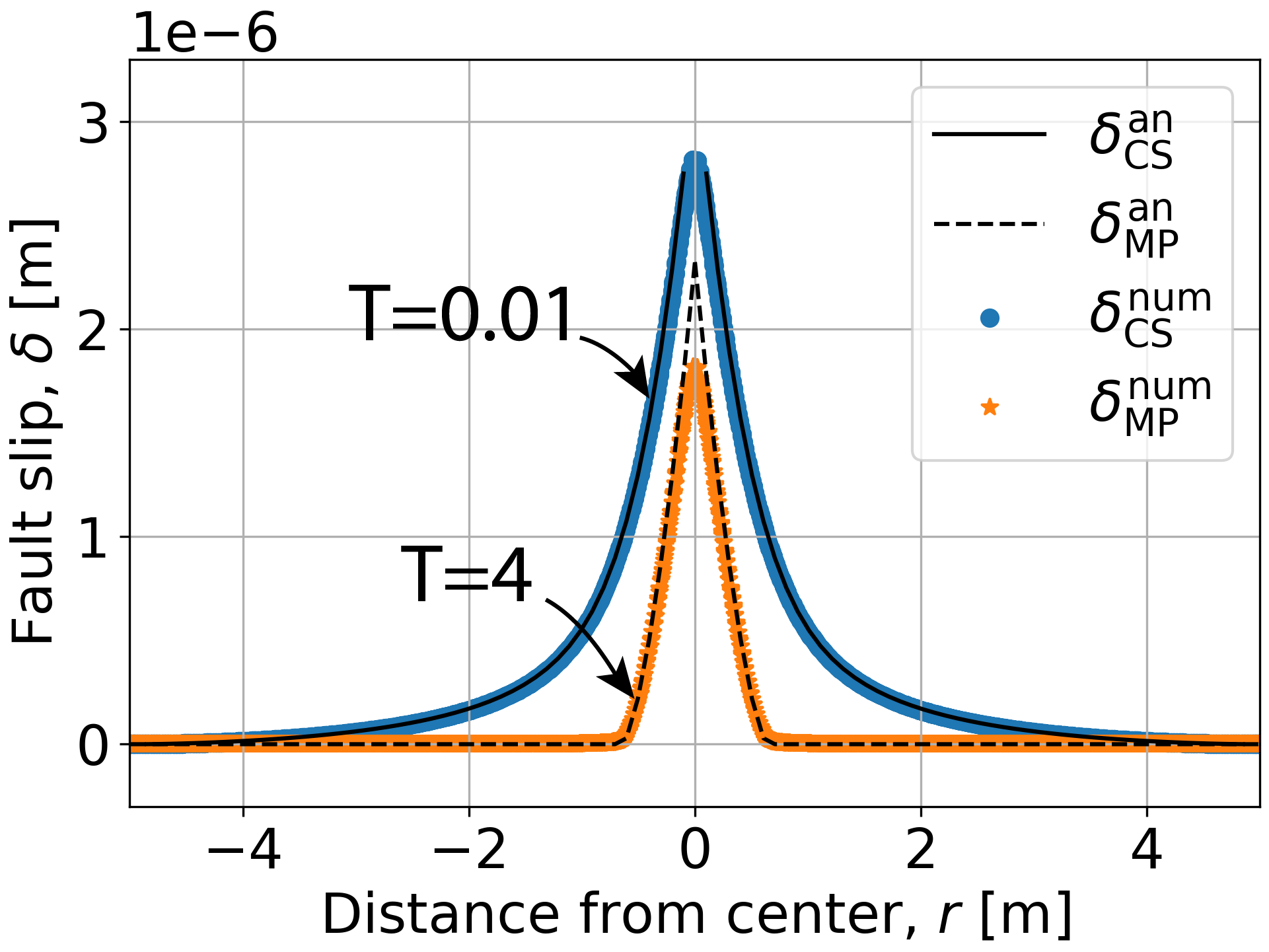}
  \end{subfigure}
  \end{center}
  \caption{Results of the critically stressed ($T=0.01$) and marginally pressurized ($T=4$) simulations at their final time steps, $t_f^{\textrm{CS}} = \qty{120}{\second}$ and $t_f^{\textrm{MP}} = \qty{5400}{\second}$, and their comparison with the analytical solutions: (\textit{left}) pressure profiles along the fracture; (\textit{right}) fracture slip along the $x$-axis. Numerical predictions and analytical solution are distinguished by superscripts $^{\mathrm{num}}$ and $^{\mathrm{an}}$ respectively, while curves representing the CS and MP limits are labelled with $_{\mathrm{CS}}$ and $_{\mathrm{MP}}$ subscripts.}
  \label{fig:injection_pressure_slip}
\end{figure}

\begin{figure} 
\centering
    \includegraphics[width=0.7\textwidth]{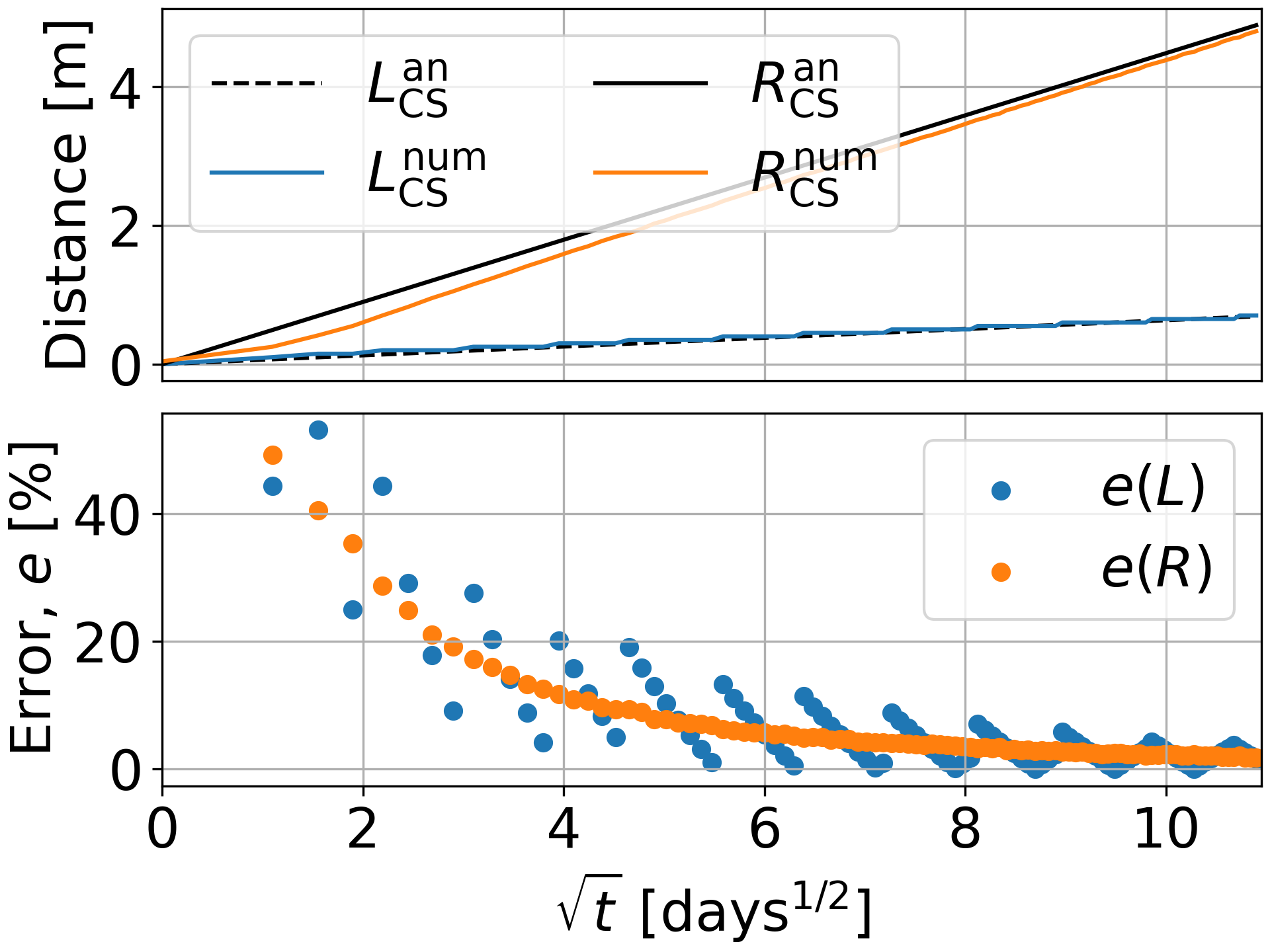}
    \caption{Numerical and analytical pressure diffusion $L(t)$ and rupture front $R(t)$ radii (top), and the evolution of their relative errors with respect to the analytical solutions of \cite{saez_three-dimensional_2022} (bottom).}
    \label{fig:injection_front}
\end{figure}

The results of the simulations and their comparison with the analytical solutions are depicted in \cref{fig:injection_pressure_slip,fig:injection_front}. On the left of \cref{fig:injection_pressure_slip}, pressure profiles at the end of MP and CS simulations perfectly align with the analytical solution, providing a benchmark for pressure diffusion problems. Notably, the MP simulation, covering a longer duration, results in a higher peak pressure and a more diffuse pressure front. The right part of \cref{fig:injection_pressure_slip} illustrates the total fracture slip, combining elastic and inelastic components, plotted along the $x$-axis. It is worth mentioning, that $\delta$ represents the slip with respect to the fracture's initial pre-injection state. When comparing the sizes of the pressure and slip fronts, consistent with theoretical predictions, the CS case features a pressure diffusion front that is narrower than the rupture front, whereas the MP case demonstrates the opposite trend.

The numerical slip closely matches the analytical solution, with only minor discrepancies in the peak slip value at the injection centre in the MP case. These differences could be further mitigated by adjusting the convergence threshold and refining the mesh. The limited slip values, measured in the order of several microns, result from the combination of a high fracture shear stiffness and a relatively small size of the domain. To mitigate the boundary effect, we chose the final time such that the maximum radii of the pressure and rupture fronts stay within $\qty{5}{\meter}$. Consequently, this restriction capped the maximum pressure and slip values. Enlarging the domain would offer the potential for higher slip magnitudes to develop.

The evolution of the pressure front $L(t)$ and the circular rupture front $R(t)$ for the critically stressed case is depicted in \cref{fig:injection_front}.
It is worth noting that the CS  case is the most computationally challenging scenario, as the discretization must capture both the pressurized region near the injection location and the rupture front which lies significantly ahead of the pore pressure front $L(t)$. 
To assess accuracy, the relative error is computed as $e(x)=|x^{\mathrm{an}}-x^{\mathrm{num}}|/x^{\mathrm{an}}$, where $x$ represents either $L(t)$ or $R(t)$. Both variables are plotted against the square root of time. The choice of the horizontal axis facilitates the understanding that the pressure diffusion and rupture fronts evolve linearly with the square root of time, consistent with theoretical expectations. Notably, both fronts closely match analytical values. The error in $L(t)$ exhibits a step-like profile due to mesh resolution, but steadily decreases, reaching below $8\%$ by the end of the simulation. Similarly, the error in $R(t)$ falls below $5\%$.

\section{Application: injection into a reservoir intersected by a permeable fault}
\label{ch:application}
\begin{figure}[]
\centering
     \includegraphics[width=0.8\textwidth]{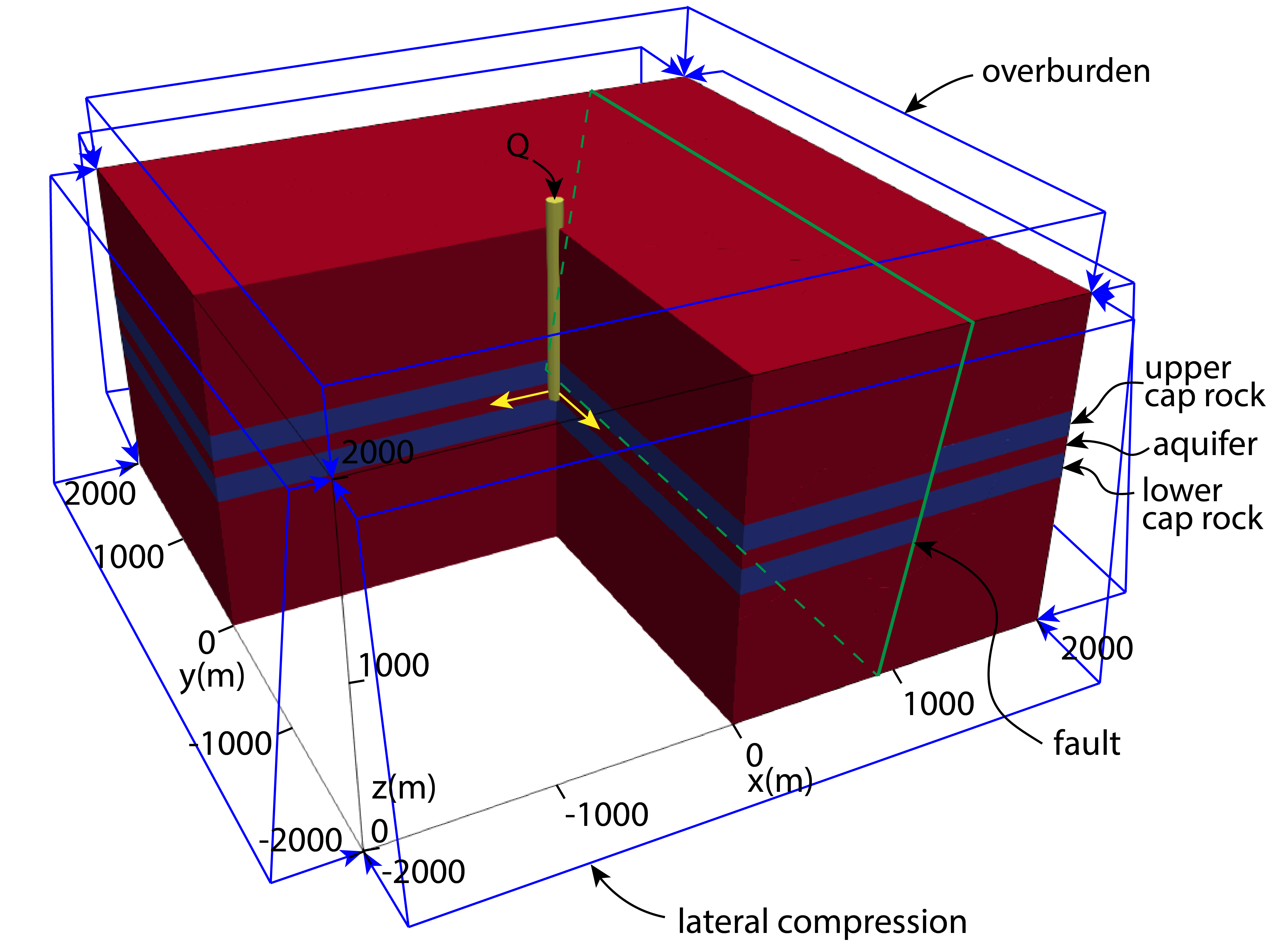}
    \caption{Model setup of $\co2$ injection into an aquifer intersected by a permeable fault}
    \label{fig:injection_aquifer}
\end{figure}

The developed solver is showcased through a simulation of $\co2$ injection into a $\qty{100}{\meter}$ thick aquifer sandwiched between two caprock layers, each $\qty{150}{\meter}$ thick, and intersected by a permeable fault. The model is illustrated in \cref{fig:injection_aquifer}, with one quarter removed for better visualization. The model size is $\qtyproduct{4000 x 4000 x 2000}{\meter}$, with the last dimension being the model height. The load applied at the top surface corresponds to the weight  due to $\qty{500}{\meter}$ of overlying rock. Minimal and maximal horizontal stresses, denoted as $\sigma_h$ and $\sigma_H$, are applied along the $x$- and $y$-axes, respectively. Lateral confinement increases linearly with depth, having gradients of $\textrm{d}\sigma_h/\textrm{d}z = \qty{11.9}{\kilo \pascal \per \meter}$ and $\textrm{d}\sigma_H/\textrm{d}z = \qty{20.4}{\kilo \pascal \per \meter}$. 
A fault, with a dip angle of $\qty{80}{\degree}$ and a strike line parallel to the $y$-axis, intersects the entire geometry. Submitted to the normal faulting regime, it has shifted the formations such that the aquifers at the hanging wall (HW) and the footwall (FW) are separated. Nevertheless, the permeable fault acts as a hydraulic bridge between two segments of the aquifer. $\co2$ is injected into the aquifer at a rate $Q=\qty{0.01}{\cubic \meter \per \second}$ along the vertical line passing through the centre of the model. The simulation is carried out until the onset of inelastic slip along the fault. This example incorporates several key aspects of geomechanical behaviour: pressure diffusion within the porous matrix and along a fault surface, poroelastic response in the matrix, and the mechanics of fault slip.

For the mechanical part, all the layers are modelled as isotropic and elastic with Young modulus $E=\qty{27.3}{\giga \pascal}$, Poisson ratio $\nu=0.15$, Biot coefficient $\alpha=0.64$, and the fault has the friction coefficient $f=0.6$, and normal and shear stiffness, $k_n=\qty{27300}{\giga \pascal}$ and $k_s=\qty{273}{\giga \pascal}$. For the fluid flow part, the aquifer, the overlying and underlying strata have permeability $k_M=1$ mD and Biot modulus $M=\qty{22.7}{\giga \pascal}$, while the caprock is tighter with $k_M=\num{1e-5}$ mD. Hydraulic properties of the fault and the fluid are taken similar to the ones in \cref{tab:material_prop}. Boundary conditions for the model are set such that the bottom surface has zero vertical displacement, and all surfaces are treated as no-flow boundaries. The tetrahedral numerical mesh contained $151,700$ degrees of freedom (three per node) and was partitioned between $72$ processors.

\begin{figure}[H]
\centering
    \begin{subfigure}{0.7\textwidth}
        \includegraphics[width=\columnwidth]{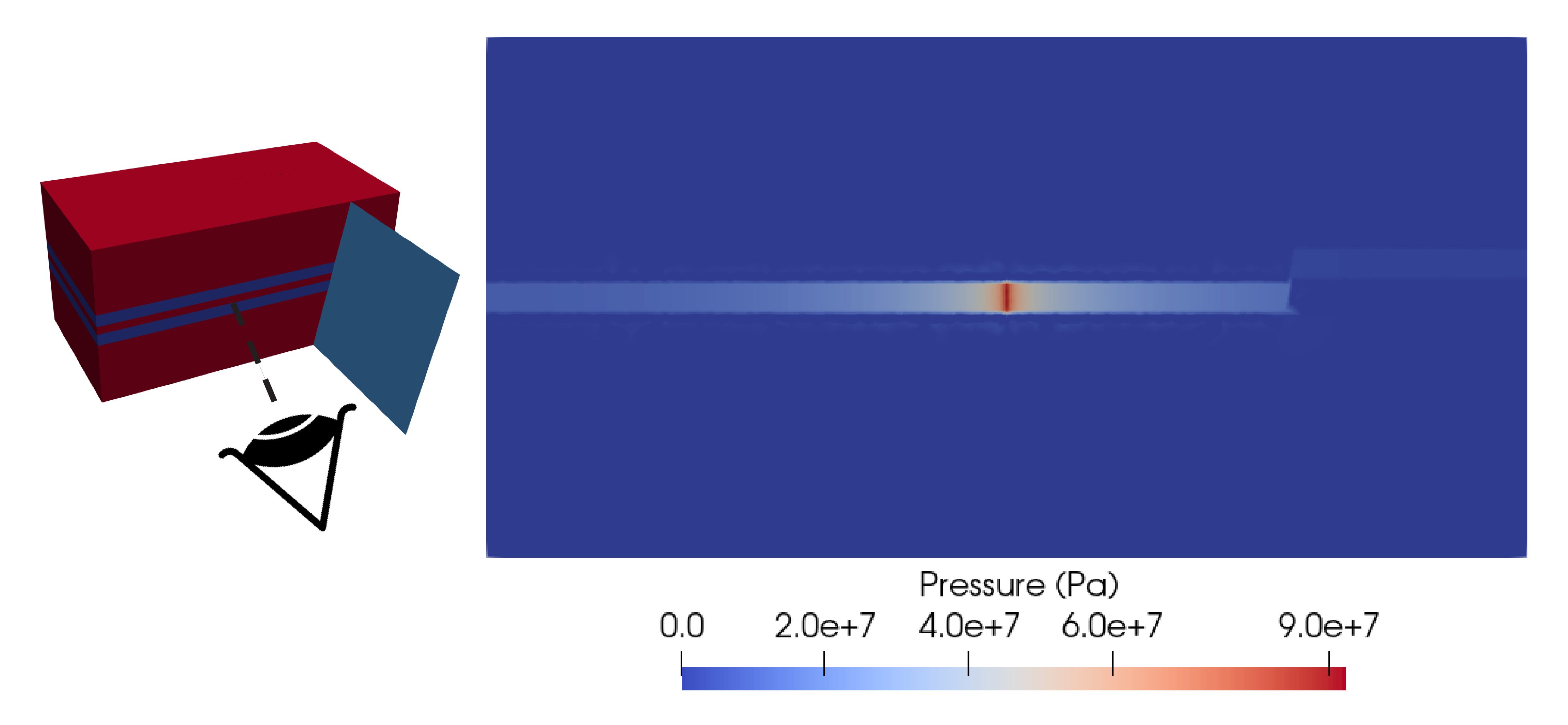}
    \end{subfigure}
    \begin{subfigure}{0.7\textwidth}
        \includegraphics[width=\columnwidth]{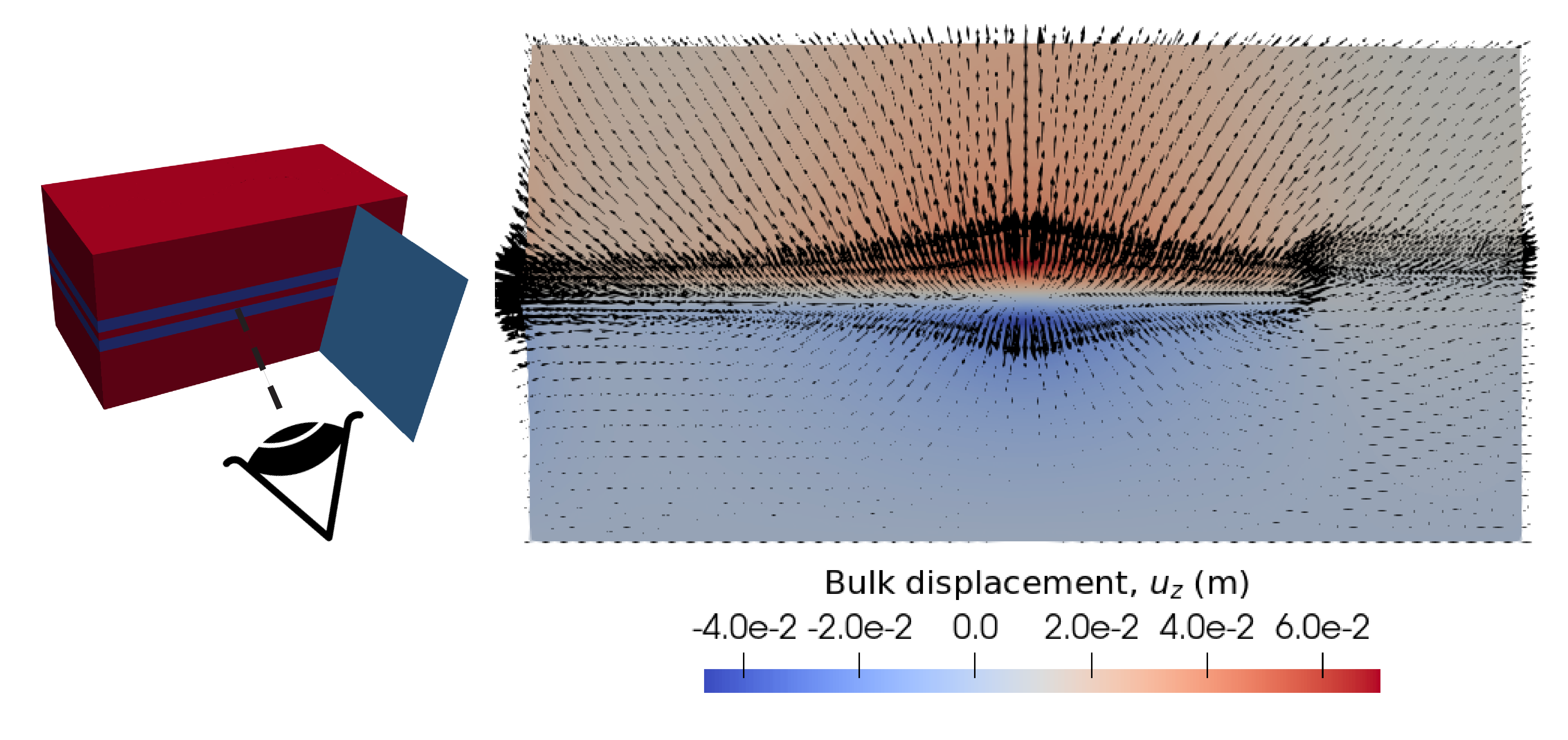}
    \end{subfigure}
    \begin{subfigure}{0.75\textwidth}
        \includegraphics[width=\columnwidth]{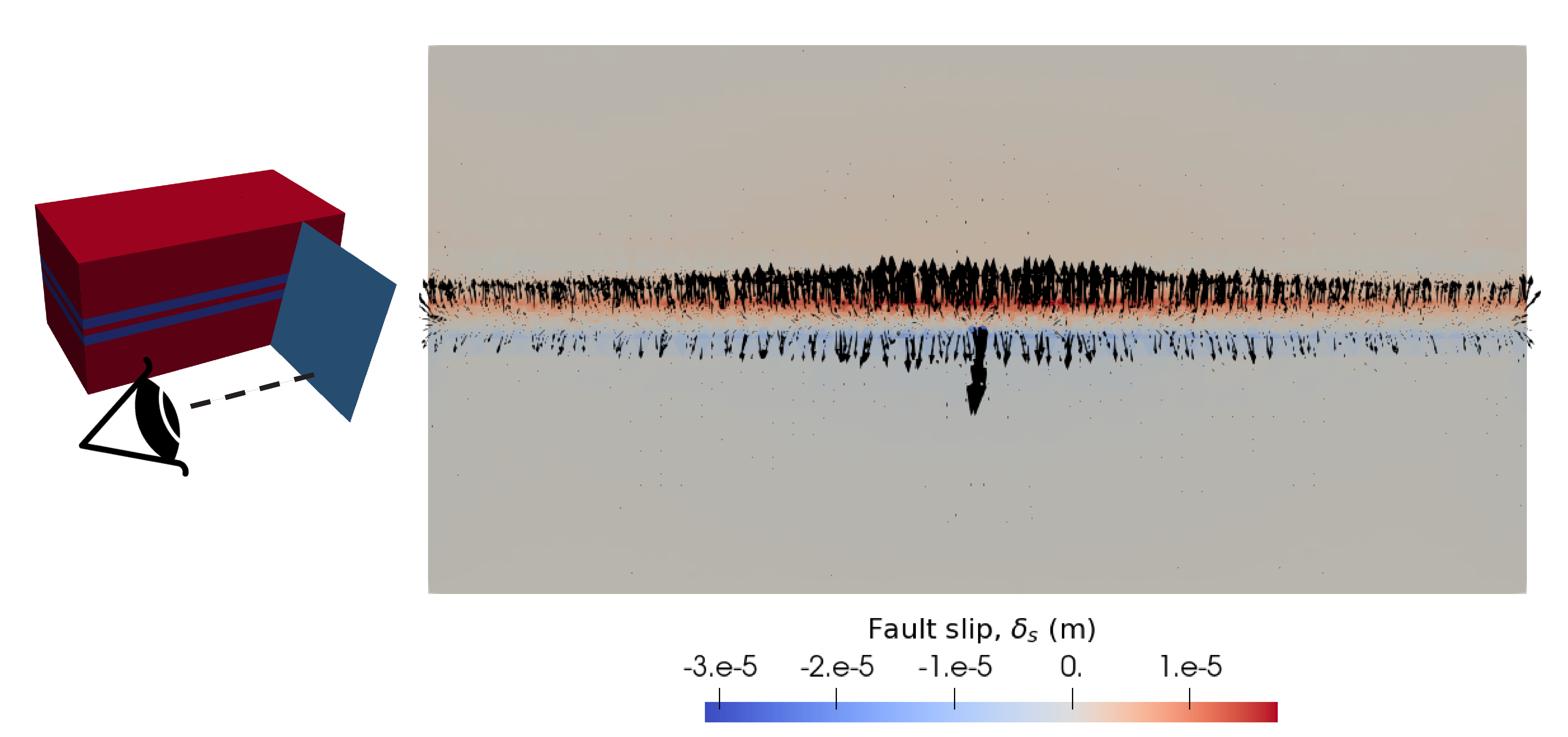}
    \end{subfigure}
    \caption{Overpressure in the matrix along the $xz$-plane (\textit{top}). Rock-mass displacement profile along the $xz$-plane (\textit{middle}). Fault slip profile (\textit{bottom}). Colours denote amplitude of the slip, while arrows show both its direction and amplitude. The arrows size is amplified to the unrealistic scale for visualization purposes.}
    \label{fig:aquifer_results}
\end{figure}

\cref{fig:aquifer_results} illustrates the distribution of pore overpressure, displacement along the $xz$ symmetry plane, and the fault slip. Top section of \cref{fig:aquifer_results} shows how pressure diffuses through the aquifer, originating at the  injection site and spreading towards the boundaries. Additionally, it demonstrates pressure transfer through the fault towards the footwall part of the aquifer. 

The observed pressure increase is accompanied by aquifer expansion, amplitude and direction of which are depicted in the middle section of \cref{fig:aquifer_results}. Although the most pronounced deformation takes place at the injection area, the expansion also extends beyond the fault. The transition from one side of the aquifer to the other occurs along the fault, leading to measurable fault slip, as depicted in the bottom part of  \cref{fig:aquifer_results}. This suggests a direct correlation between the aquifer pressure dynamics and fault mechanics.
The arrows in the bottom part of \cref{fig:aquifer_results} indicate the amplified magnitude and direction of the slip, while the colour represents the amplitude of the slip projection along the vertical axis. The fault slip comprises both elastic and inelastic component. While most of the arrows correspond to the elastic slip, the largest arrow, located in the centre, indicates the point where the inelastic slip begins, suggesting a potential origin of instability. The alignment of this large arrow with the fault dip line suggests the direction of the potential slip event.

\begin{figure}[]
\centering
    \includegraphics[width=0.8\textwidth]{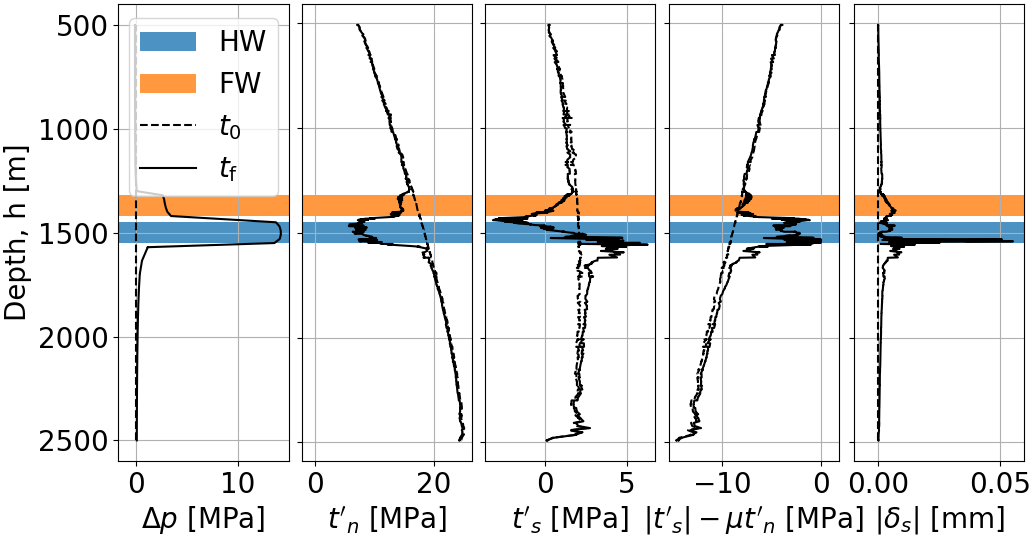}
    \caption{Hydro-mechanical fields plotted along the dip line within the $xz$-plane.}
    \label{fig:fault_figures}
\end{figure}

\Cref{fig:fault_figures} illustrates a collection of hydro-mechanical fields projected along the dip line passing through the centre of the fault. The plotted fields include overpressure $\Delta p$, effective normal traction $t'_n$, projection of frictional force on the fault dip line $t'_s$, yield function $\mathcal{F}(t') =|t'_s| - \mu t'_n$, and the total slip norm $|\delta_s|$. The vertical axis represents the absolute depth of a point, while the blue- and orange-shaded regions indicate the depths where the hanging wall (HW) and footwall (FW) parts of the aquifer contact the fault.  The dashed lines represent the pre-injection state, while the solid line reflect the state at the end of simulation.

Initially, pressure increases in the HW section and later in the FW section, leading to a reduction in effective traction $t'_n$. Simultaneously, aquifer expansion results in a rise in frictional force along the fault, manifesting as two opposite peaks in the $t'_s$ profile. The combination of decreasing $t'_n$ and increasing $t'_s$ leads to a rise in the yield function $\mathcal{F}(t')$, eventually reaching zero. At this point, inelastic slip is triggered, as evidenced by a significant peak in the slip profile. The reason this specific central point of the fault is the first to slip is due to the combined decrease of frictional resistance (effective normal stress) and increase of shear stress. 

\begin{figure}[H]
\centering
    \begin{subfigure}{0.3\textwidth}
        \includegraphics[width=\columnwidth]{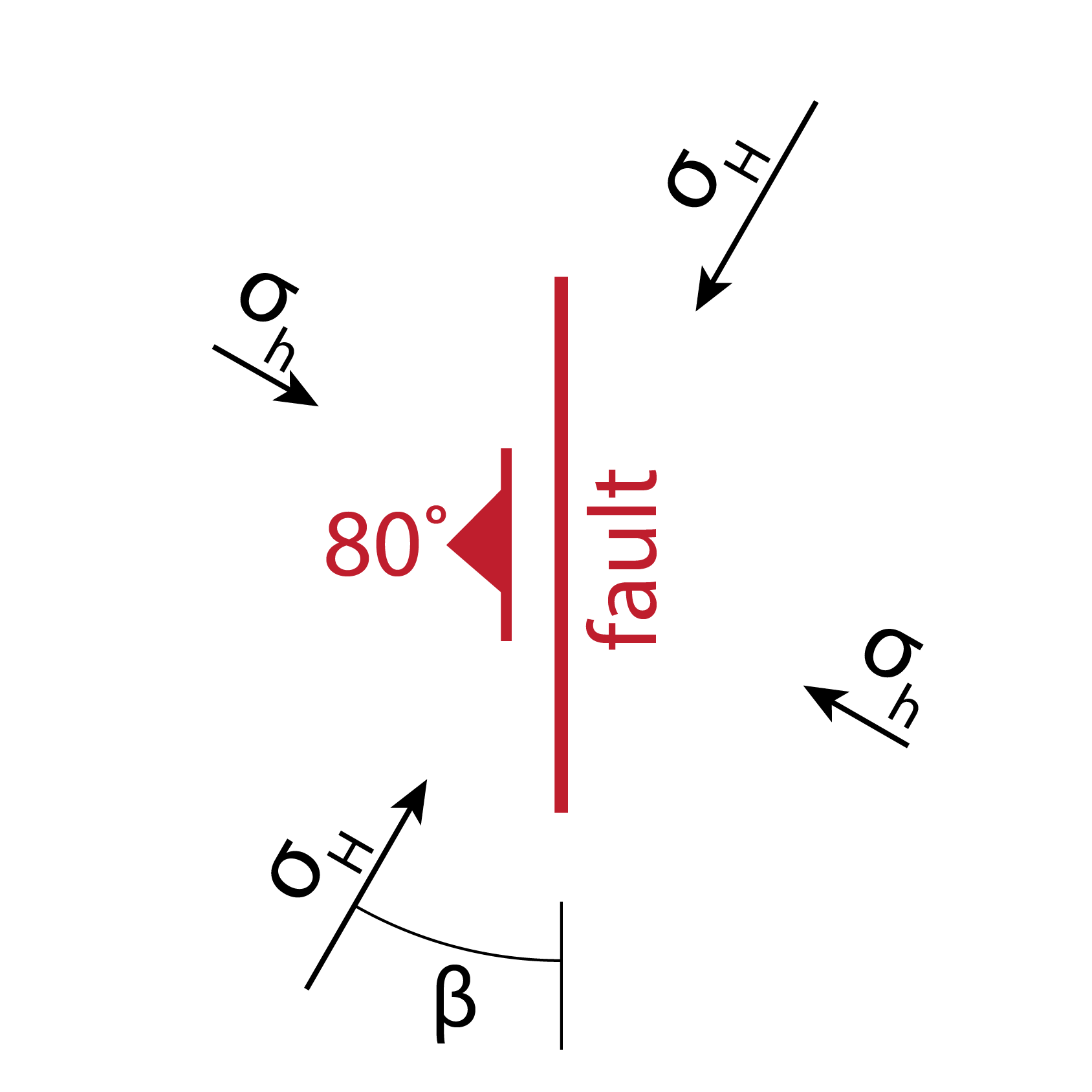}
    \end{subfigure}
    \begin{subfigure}{0.4\textwidth}
        \includegraphics[width=\columnwidth]{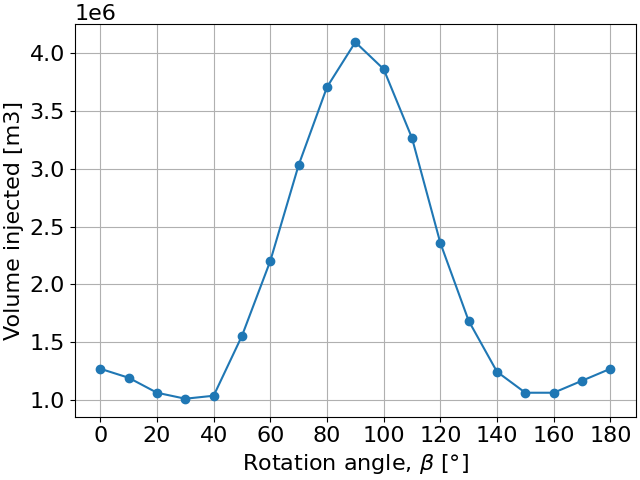}
    \end{subfigure}
    \begin{subfigure}{0.75\textwidth}
        \includegraphics[width=\columnwidth]{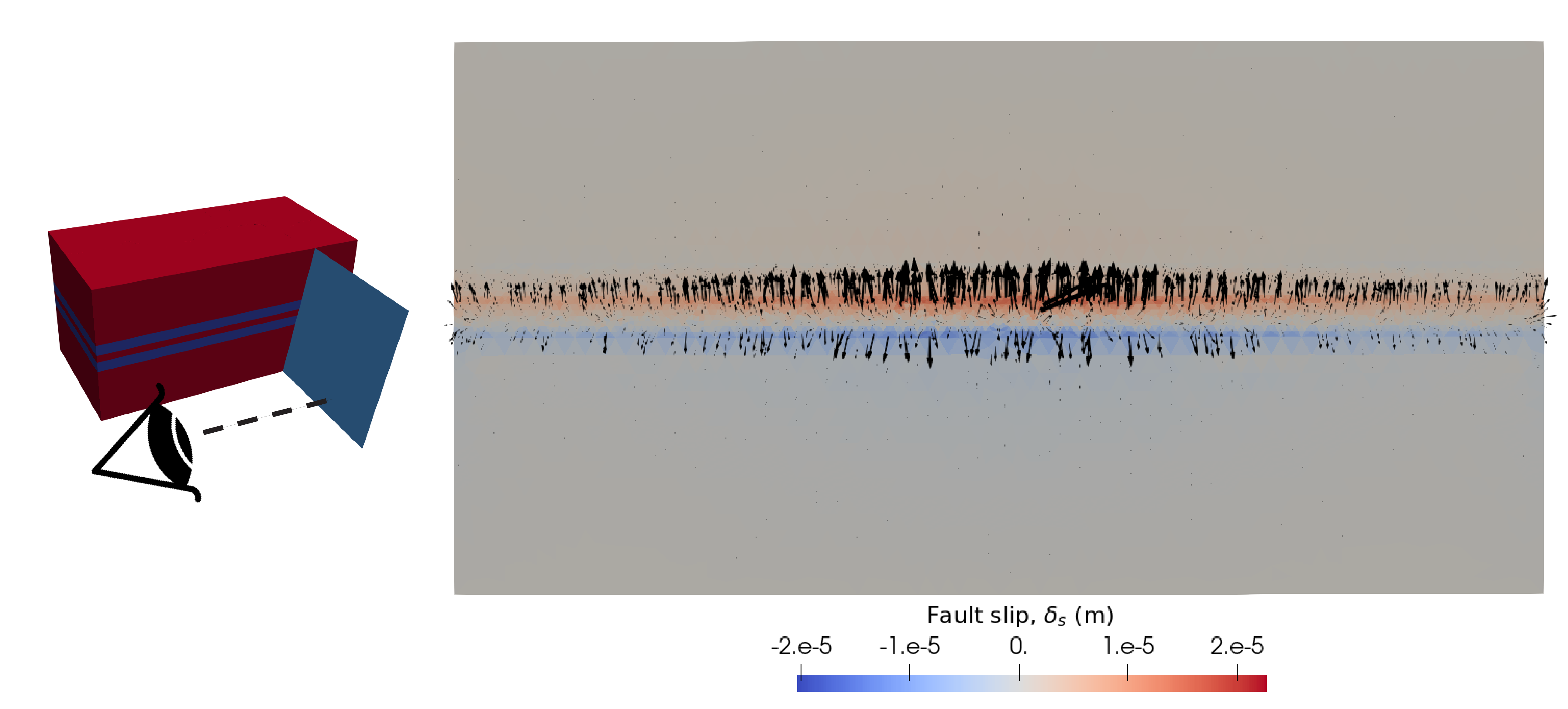}
    \end{subfigure}
    \caption{Angle of rotation of horizontal stresses relative to the fault (\textit{top left}). $\co2$ volume injected before the onset of inelastic slip against rotation angle $\beta$ (\textit{top right}). Slip profile for rotation angle $\beta=\qty{30}{\degree}$ (\textit{bottom}).}
    \label{fig:fault_orientation}
    \end{figure}

The fault slip profile and the evolution of hydro-mechanical fields are specific to a fault orientation where horizontal stresses are aligned normally to the fault strike line. To explore variations in fault behaviour under different stress conditions, we conducted similar simulations with rotated horizontal stresses relative to the fault strike line by angle $\beta$, as depicted in the top left part of \cref{fig:fault_orientation}.
The time before the onset of inelastic slip depends on the rotation angle $\beta$. The top right section of \Cref{fig:fault_orientation} presents the volume of injected $\co2$ required to trigger slip as a function of this rotation angle. The data reveal a distinct bell-shaped curve centred at $\qty{90}{\degree}$, indicating that this orientation, where the maximum horizontal stress $\sigma_H$ is nearly perpendicular to the fault, is the most favourable orientation to avoid inducing slip.

Stress orientations that promote fault slip are primarily around $\qty{30}{\degree}$ and $\qty{150}{\degree}$, at which the maximum initial shear stress along the fault plane occurs. A slip profile for $\beta=\qty{30}{\degree}$ is depicted in the bottom part of \cref{fig:fault_orientation}, with inelastic slip indicated by the inclined large arrows in the centre of the fault. Unlike the default configuration at $\beta=\qty{0}{\degree}$, where the inelastic slip occurs vertically, here the inelastic slip is close to horizontal, characteristic of a strike-slip regime. 

This pattern confirms that when the fault plane is at an angle of approximately $\qty{30}{\degree}$ to $\qty{45}{\degree}$, the risk of slip is higher, requiring smaller amounts of $\co2$ to be injected. Conversely, when the principal stresses deviate from this orientation, more $\co2$ can be injected before slip occurs, indicating enhanced fault stability. Understanding this relationship is crucial for managing risks in subsurface operations, such as $\co2$ sequestration and other geological engineering applications.

\section{Concluding remarks}
We have developed a new geomechanical model designed to simulate geological fluid injection and extraction. The model offers several key advantages:

\begin{enumerate}
    \item It allows for two-way coupling between displacement and pressure fields.
    \item Through an iterative partitioned procedure for the simultaneous solution of the tangent system of coupled geomechanical equations, it ensures unconditional stability and computational efficiency of solution.
    \item It represents planar rock discontinuities as embedded surfaces in three-dimensional domain, enabling sharp discontinuities in displacement, essential for fault slip. 
    \item The two-node flux approximation in fractures facilitates the definition of independent fracture pressure and supports different longitudinal and transversal fracture permeabilities.
    \item Coulomb friction along the fracture allows for the modelling of fault reactivation, but it also introduces non-linearity in the mechanical response of the model.
    \item The model is designed for high-performance computing using the MPI library.  
\end{enumerate}

This is the first time, in the literature, that the iterative partitioned conjugate gradient approach introduced by \cite{prevost_partitioned_1997} is applied to problems involving explicitly defined discontinuities, such as fractures and faults. The proposed finite-element implementation maintains the computational efficiency and modularity that \citeauthor{prevost_partitioned_1997} demonstrated for the continuous version of the solver.

Although we assumed constant matrix permeability and fracture transmissibility in the presented examples, variations in these parameters can become significant when there are large changes in stress and normal traction. Integrating these dependencies into the model and examining its convergence properties will be a priority of future research. 

An additional significant advancement would be to incorporate multiple phases in the fluid flow sub-problem, like gas and liquid, allowing for more accurate modelling of the pressure profile. Given the modularity inherent to the solution framework, this could be achieved by plugging-in an external multiphase flow solver in the block conjugate gradient scheme.

The proposed geomechanical solver is a versatile tool for examining a wide range of application, including geotechnical engineering, tunnelling, $\co2$ sequestration, water, hydrocarbons, and geothermal energy extraction, as well as related phenomena like fault reactivation, hydraulic fracturing, and surface subsidence or uplift. It benefits from the advanced numerical features of the open-source FEM code Akantu, including dynamic insertion of fracture cohesive elements, and more intricate friction laws.

The solver accuracy in relation to shear fracture slip has been verified against novel analytical solutions of the propagation of frictional rupture induced by fluid flow. We re-emphasize the importance of this test in the development of a robust solver for this class of non-linear hydro-mechanical problem. Such in-depth comparison is necessary to advance further the accuracy and robustness of numerical methods for the simulation of moving discontinuities in porous media.

The open-source code for the geomechanical solver, along with various tutorials and benchmarks, is publicly available\footnote{\url{https://gitlab.com/emil.gallyamov/akantu-geomechanical-solver}}. Additionally, visualization of $\co2$ injection into a faulted aquifer presented in \cref{ch:application} is also accessible online\footnote{\url{https://renkulab.io/projects/phamba/geology-data-visualization}}. Full dataset with simulation results of injection are available at the permalink 10.5281/zenodo.13710229.

\section*{Acknowledgements}
The results were obtained within the OSGEOCGS project (An Open-Source platform for Geomechanical assessment of $\co2$ Geological Storage) funded by an EPFL ENAC Interdisciplinary Cluster Grant (2022-2024).

\bibliography{references}

\newpage
\appendix
\section{Numerical implementation of coupling operators of $\boldsymbol{Ap}$-type for solid elements}
\label{app:coupling}
In this section, we will explicitly define coupling operators for solid and fracture elements in a form which is ready to be implemented in a standard finite element code. To do so, we will first establish general notations and fundamental operators in the context of the finite element method (FEM). For simplicity, we will focus on a 2D problem using linear triangular elements for the solid domain and linear duplicated segments for the fracture. The transformation of the derived operators to 3D is straightforward and is left as an exercise for the reader. Symmetric tensors will be presented in Voigt notation.

By employing a nodal-based finite element approximation, the displacement at a point $\xi$ within the solid region is given by
\begin{equation}
        \{u(\xi)\} = [N_u(\xi)]\{u\},   
\end{equation}
where $[N_u(\xi)]$ is the matrix of shape functions values $N$ at a point $\xi$, and $\{d\}$ is the vector of nodal displacements. We expand these two terms as follows:
\begin{equation}
\begin{split}
    &[N_u(\xi)]=
    \begin{bmatrix}
        N_1(\xi) & 0 & N_2(\xi) & 0 & N_3(\xi) & 0 \\
        0 & N_1(\xi) & 0 & N_2(\xi) & 0 & N_3(\xi)
    \end{bmatrix},\\
    &\{u\}=
    \begin{Bmatrix}
        h_1 & v_1 & h_2 & v_2 & h_3 & v_3
    \end{Bmatrix}^T,
\end{split}
\end{equation}
where $h$ and $v$ are the horizontal and vertical components of the nodal displacements, respectively.

The piecewise constant strain is computed as:
\begin{equation}
        \{\varepsilon\} =
        \begin{Bmatrix}
            \varepsilon_x\\
            \varepsilon_y\\
            \tau_{xy}
        \end{Bmatrix}
        =[B_u]\{u\},   
\end{equation}
where $[B_u]$ is the matrix of shape function derivatives, which can be expressed as:
\begin{equation}
    [B_u]=
    \begin{bmatrix}
        \frac{\partial N_1}{\partial x} & 0 & \frac{\partial N_2}{\partial x} & 0 & \frac{\partial N_3}{\partial x} & 0 \\
        0 & \frac{\partial N_1}{\partial y} & 0 & \frac{\partial N_2}{\partial y} & 0 & \frac{\partial N_3}{\partial y}\\
        \frac{\partial N_1}{\partial y} & \frac{\partial N_1}{\partial x} & \frac{\partial N_2}{\partial y} & \frac{\partial N_2}{\partial x} & \frac{\partial N_3}{\partial y} & \frac{\partial N_3}{\partial x}
    \end{bmatrix}.\\
    \end{equation}

Similarly, operators are established to interpolate the pressure at a point $\xi$ and to compute pressure gradients $\nabla p$:
\begin{equation}
\begin{split}
    & p(\xi) = [N_p(\xi)]\{p\},\\
    & {\nabla p} = [B_p]\{p\},\\
    &[N_p(\xi)]=
    \begin{bmatrix}
        N_1(\xi) & N_2(\xi) & N_3(\xi)
    \end{bmatrix},\\
    &[B_p]=
    \begin{bmatrix}
        \frac{\partial N_1}{\partial x} & \frac{\partial N_2}{\partial x} & \frac{\partial N_3}{\partial x} \\
        \frac{\partial N_1}{\partial y} & \frac{\partial N_2}{\partial y} & \frac{\partial N_3}{\partial y}
    \end{bmatrix},\\
    &\{p\}=
    \begin{Bmatrix}
        p_1 & p_2 & p_3
    \end{Bmatrix}^T.
\end{split}
\end{equation}

Having defined the basic operators, we can construct the second term in \cref{eq:discrete_balance_full}, expressed as:
\begin{equation}
    \int_{\Omega_M} \biot \{\varepsilon(\testv)\}^T \{I\} p_{\textrm{IP}} \ d\Omega,
\end{equation}
where $\{I\} = \begin{Bmatrix}
    1 & 1 & 0
\end{Bmatrix}^T$ is the identity tensor, and $p_{\textrm{IP}}$ represents the pressure interpolated at the integration points. After conducting the standard finite element operation of eliminating the nodal values of the test function $\{\testv\}$, we obtain the product $\boldsymbol{A}^{\textrm{m}}_{\textrm{pu}}\boldsymbol{p}$, which couples the fluid pressure to the matrix deformation:
\begin{equation}
\label{eq:Apdp}
\begin{split}
\boldsymbol{A}^{\textrm{m}}_{\textrm{pu}} \boldsymbol{p} =& \int_{\Omega_M} \biot [B_u]^T \{I\} [N_p]\{p\} \ d\Omega \\
=&
\int_{\Omega_M} \biot
\begin{bmatrix}
    \frac{\partial N_1}{\partial x} & 0 & \frac{\partial N_1}{\partial y}\\  
    0 & \frac{\partial N_1}{\partial y} & \frac{\partial N_1}{\partial x}\\
    \frac{\partial N_2}{\partial x} & 0 & \frac{\partial N_2}{\partial y}\\
    0 & \frac{\partial N_2}{\partial y} & \frac{\partial N_2}{\partial x} \\
    \frac{\partial N_3}{\partial x} & 0 & \frac{\partial N_3}{\partial y}\\
    0 & \frac{\partial N_3}{\partial y} & \frac{\partial N_3}{\partial x}
\end{bmatrix}
\begin{Bmatrix}
    1 \\ 
    1 \\ 
    0
\end{Bmatrix}
\begin{bmatrix}
    N_1 & N_2 & N_3
\end{bmatrix}
\begin{Bmatrix}
    p_1 \\ 
    p_2 \\ 
    p_3
\end{Bmatrix}.
\end{split}
\end{equation}

The first term of \cref{eq:discrete_mass}, which couples the displacement field to the pressure field, is rewritten assuming the first-order time integration and using Voigt notation:
\begin{equation}
    \int_{\Omega_M} r_{\textrm{IP}} \biot \{I\}^T \{\varepsilon(\Delta \disp)\} \ \text{d}\Omega / \Delta t,   
\end{equation}
where $r_{\textrm{IP}}$ denotes the scalar test function $r$ evaluated at the integration points, $\Delta \disp$ is the displacement increment within a time step,  $\varepsilon(\Delta \disp)$ is the resulting strain increment, and $\Delta t$ is the time step size. After performing finite element manipulations similar to those above and omitting the $\Delta t$ term, we recover the second coupling product: 
\begin{equation}
\label{eq:Adpd}
\begin{split}
    &\boldsymbol{A}^{\textrm{m}}_{\textrm{up}} \boldsymbol{\Delta u} = \int_{\Omega_M} \biot [N_p]^T \{I\}^T [B_u]\{\Delta u\} \ d\Omega \\
&=
\int_{\Omega_M} \biot
\begin{bmatrix}
    N_1\\
    N_2\\
    N_3
\end{bmatrix}
\begin{Bmatrix}
    1 & 1 & 0
\end{Bmatrix}
\begin{bmatrix}
    \frac{\partial N_1}{\partial x} & 0 & \frac{\partial N_2}{\partial x} & 0 & \frac{\partial N_3}{\partial x} & 0 \\
    0 & \frac{\partial N_1}{\partial y} & 0 & \frac{\partial N_2}{\partial y} & 0 & \frac{\partial N_3}{\partial y}\\
    \frac{\partial N_1}{\partial y} & \frac{\partial N_1}{\partial x} & \frac{\partial N_2}{\partial y} & \frac{\partial N_2}{\partial x} & \frac{\partial N_3}{\partial y} & \frac{\partial N_3}{\partial x}
\end{bmatrix}
\begin{Bmatrix}
    h_1 \\ 
    v_1 \\ 
    h_2 \\ 
    v_2 \\ 
    h_3 \\ 
    v_3 
\end{Bmatrix}.
\end{split}
\end{equation}
In \cref{eq:Apdp,eq:Adpd}, the shape functions $[N_p]$ are evaluated at the integration points.

\section{Numerical implementation of coupling operators of $\boldsymbol{A}\boldsymbol{p}$-type for fracture elements}
\label{app:coupling_fracture}

\begin{figure}[htbp]
  \centering
  \includegraphics[width=0.3\columnwidth]{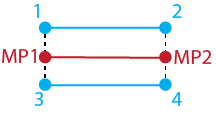}
  \caption{Topology of a linear 2D fracture element}
  \label{fig:cohesive_2D}
\end{figure}

Since the topology of the interface elements used to discretize fractures differs from that of solid elements, the corresponding finite element operators are defined separately. The topology of a linear 2D interface element is illustrated in \cref{fig:cohesive_2D}
The mid-points shown in the figure are fictitious and are used for illustrative purposes only.

The opening of an interface element at a point $\psi$ is calculated based on the nodal displacements:
\begin{equation}
    \{\delta (\psi)\}=
    \begin{Bmatrix}
        \delta_x\\
        \delta_y
    \end{Bmatrix}
    = [N_u^f(\psi)]
    \begin{bmatrix}
        I & -I
    \end{bmatrix}
    \{u\},
\end{equation}
where $\{u\}=\begin{Bmatrix}
    h_1 & v_1 & h_2 & v_2 & h_3 & v_3 & h_4 & v_4
\end{Bmatrix}^T$ represents the nodal displacement vector, $I$ is the $4$-by-$4$ identity tensor, and
$[N_u^f(\psi)]$ is the matrix containing values of the shape functions for the fracture elements evaluated at the point $\psi$. It can be explicitly expressed as:
\begin{equation}
    [N_u^f(\psi)] =
    \begin{bmatrix}
        N_{\textrm{MP1}}(\psi) & 0 & N_{\textrm{MP2}}(\psi) & 0 \\
        0 & N_{\textrm{MP1}}(\psi) & 0 & N_{\textrm{MP2}}(\psi)
    \end{bmatrix}.
\end{equation}

The normal opening is calculated by multiplying the opening vector with the normal vector $\mathbf{n}$:
\begin{equation}
    \delta_n = \{n\}^T \{\delta \}  = 
    \begin{Bmatrix}
        n_{x} & n_{y}
    \end{Bmatrix}
    \begin{Bmatrix}
        \delta_x \\
        \delta_y
    \end{Bmatrix}
\end{equation}

Similarly, the average pressure at the point $\psi$ is computed as:
\begin{equation}
     \av{p(\psi)} = [N_p^f(\psi)] \frac{1}{2} 
    \begin{bmatrix}
        I_p & I_p
    \end{bmatrix}
    \{p\},
\end{equation}
where
\begin{equation}
\begin{split}
    \{p\}=&
    \begin{Bmatrix}
        p_1 & p_2 & p_3 & p_4
    \end{Bmatrix},\\
    [N_p^f(\psi)]=&
    \begin{bmatrix}
        N_{\textrm{MP1}}(\psi) & N_{\textrm{MP1}}(\psi)
    \end{bmatrix},
\end{split}
\end{equation}
and $I_p$ is an identity matrix whose order equals the number of mid-points, which, in the case of linear 2D interface elements, is $2$.

We now turn our attention to the first term in the weak form of the fluid volume conservation equation within a fracture, presented in \cref{eq:discrete_mass_fracture}:
\begin{equation}
    \int_{\Gamma_{MF}} \av{r_{\textrm{IP}}} \frac{\partial w_{\textrm{IP}}}{\partial t} \text{d} \Gamma,
\end{equation}
where $\av{r_{\textrm{IP}}}$ represents the averaged values of the scalar test function $\{r\}=
\begin{Bmatrix}
      r_1 & r_2 & r_3 & r_4
\end{Bmatrix}^T$ interpolated at the integration points, and $w_{\textrm{IP}}$ denotes the hydraulic opening at the integration points. The above term couples the velocity of normal fracture opening $\partial w /\partial t$ to pressure. After applying first-order time integration, assuming that the increment of hydraulic aperture $\Delta w$ is equal to the increment of normal fracture opening $\Delta \delta_n$, and substituting the finite element operators defined earlier, while eliminating the test function, we derive the displacement-to-pressure coupling product within the fracture:
\begin{equation}
\label{eq:afdpd}
    \boldsymbol{A}^{\textrm{f}}_{\textrm{up}} \boldsymbol{\Delta u} = \frac{1}{2} \int_{\Gamma_{MF}}
    \begin{bmatrix}
        I_p & I_p
    \end{bmatrix}^T
     [N_p^f]^T \{n\}^T [N_u^f] 
    \begin{bmatrix}
        I & -I
    \end{bmatrix}
    \{\Delta u \} \ \text{d} \Gamma.
\end{equation}

Finally, the pressure-to-displacement coupling term, which converts fluid pressure into a mechanical force acting on the fracture lips, is found in \cref{eq:discrete_balance_full}:
\begin{equation}
    \int_{\Gamma^+_{MF}\cup \Gamma^-_{MF}} \av{\testv_{\textrm{IP}}}^T \cdot \av{p_{\textrm{IP}}} \{n^{\pm}\}  \ d\Gamma.
\end{equation}
Substituting in the finite element operators and eliminating the test function, we find the algorithmic form of the last coupling product:
\begin{equation}
\label{eq:afpdp}
    \boldsymbol{A}^{\textrm{f}}_{\textrm{pu}} \boldsymbol{p}= 
    \frac{1}{4} \int_{\Gamma^+_{MF}\cup \Gamma^-_{MF}} 
    \begin{bmatrix}
        I & I
    \end{bmatrix}^T
    [N_u^f]^T [N_p^f]
    \begin{bmatrix}
        I_p & I_p
    \end{bmatrix}
    \{n^{\pm}\}    \{p\}  \ d\Gamma.
\end{equation}
In \cref{eq:afdpd,eq:afpdp}, the shape functions $[N_p^f]$ and $[N_u^f]$ are evaluated at the integration points of interface elements.

As outlined in the original work by \cite{prevost_partitioned_1997}, the global coupling matrices do not need to be directly constructed. Instead, $\boldsymbol{Ap}$-type products can be calculated at the level of integration points, integrated over the corresponding volume or area, and consequently aggregated into the relevant global vectors.

\end{document}